\documentclass[journal,twocolumn]{IEEEtran}
\usepackage{CJKutf8}
\usepackage{array}
\usepackage{booktabs} %调整表格线与上下内容的间隔
\usepackage{multirow}
\usepackage{bm}
\usepackage{mathrsfs}
\usepackage{cite}
\usepackage{amsmath}
\usepackage{amsthm}
\usepackage{amssymb,amsfonts}
\usepackage{graphicx}
\usepackage{textcomp}
\usepackage{xcolor}
\usepackage[T1]{fontenc}
\usepackage[utf8]{inputenc}
\usepackage[font=small,tableposition=top]{caption}
\usepackage{booktabs}
\usepackage{url}
\usepackage{algorithm}
\usepackage{algpseudocode}
\usepackage{verbatim}

\usepackage{soul}
\usepackage{caption}
\usepackage{tabularx}
\usepackage{indentfirst}

\usepackage{nomencl}%%%%术语表
\makenomenclature

\usepackage{subfig}

\usepackage{CJK}
\usepackage{type1cm}
\usepackage{times}
\usepackage[marginal]{footmisc}
\usepackage{stfloats}%表格不会跳到下一页

\def\BibTeX{{\rm B\kern-.05em{\sc i\kern-.025em b}\kern-.08em
		T\kern-.1667em\lower.7ex\hbox{E}\kern-.125emX}}
\ifCLASSINFOpdf
\else
 \fi

\begin{document}
\title{Joint Trading and Scheduling among Coupled Carbon-Electricity-Heat-Gas Industrial Clusters}

%%%%%%%%%%%%%%作者!!!!!!!!!!!!!!!!!! Distributed Stochastic Gradient Optimization of Multi-Energy Management in an Industrial Park
%\author{-
%	\thanks{This work was supported in part by the National Key Research and Development Program of China (2016YFB0901900), in part by the NSF of
%		China (61922058, 61603251, and	61731012), and NSF of Shanghai Municipality (18ZR1419900). }
%    \thanks{The authors are with the Department of Automation, Shanghai Jiao Tong University, and also with the Key Laboratory of System Control and Information Processing, Ministry of Education of China, Shanghai 200240, China (e-mail: dafeng.zhu@sjtu.edu.cn; bo.yang@sjtu.edu.cn).}}

%\markboth{Journal of \LaTeX\ Class Files,~Vol.~14, No.~8, August~2015}%
%{Shell \MakeLowercase{\textit{et al.}}: Bare Demo of IEEEtran.cls for IEEE Journals}

%%%%%%%%%%%%%%%%%%曹师姐作者模板

\author{Dafeng~Zhu,
	\IEEEmembership{Member,~IEEE,}
	Bo~Yang,
	\IEEEmembership{Senior~Member,~IEEE,}
	Yu~Wu,
	\IEEEmembership{Student~Member,~IEEE,}
	Haoran~Deng,
	\IEEEmembership{\\Student~Member,~IEEE,}
	Zhaoyang~Dong,
	\IEEEmembership{Fellow,~IEEE,}
	%Chengbin~Ma,
	%~\IEEEmembership{Senior~Member,~IEEE,}
	%\\Zhaojian~Wang,
	%~\IEEEmembership{Member,~IEEE,}
	%Shanying~Zhu,
	%~\IEEEmembership{Member,~IEEE,}
	Kai~Ma,
	\IEEEmembership{Member,~IEEE}
	%Cailian~Chen,
	%\IEEEmembership{Member,~IEEE}
	and~Xinping~Guan,
	\IEEEmembership{Fellow,~IEEE}
	%\thanks{\indent This work was supported in part by the National Natural Science Foundation of China ( 61731012, 62273237, 62025305, 61933009, 92167205).  \emph{(Corresponding author: Dafeng Zhu.)}}
	%\thanks{\indent Bo Yang, Dafeng Zhu, and Xinping Guan are with the Department of Automation, Shanghai Jiao Tong University, Shanghai 200240, China; Key Laboratory of System Control and Information Processing, Ministry of Education of China, Shanghai 200240, China; Shanghai Engineering Research Center of Intelligent Control and Management, Shanghai 200240, China (e-mail: bo.yang@sjtu.edu.cn; dafeng.zhu@sjtu.edu.cn; xpguan@sjtu.edu.cn).\\
	\thanks{\indent Dafeng Zhu, Bo Yang, Yu Wu, Haoran Deng and Xinping Guan are with the Department of Automation, Shanghai Jiao Tong University, Shanghai 200240, China (e-mail: dafeng.zhu@sjtu.edu.cn; bo.yang@sjtu.edu.cn; 5wuuy5@sjtu.edu.cn; denghaoran@sjtu.edu.cn; xpguan@sjtu.edu.cn).\\
	\indent Zhaoyang Dong  is with the School of Electrical and Electronic Engineering, Nanyang Technological University, Singapore 639798 (e-mail: zydong@ieee.org).\\
	%\indent C. Ma is with the University of Michigan-Shanghai Jiao Tong University Joint Institute, Shanghai Jiao Tong University, Shanghai 200240, China (chbma15@gmail.com).\\
	\indent Kai Ma is with the School of Electrical Engineering, Yanshan University, Qinhuangdao 066004, China (e-mail: kma@ysu.edu.cn).}}
	%\indent Kai Ma is with the School of Electrical Engineering, Yanshan University, Qinhuangdao 066004, China (e-mail: kma@ysu.edu.cn).
	%\indent Kai Ma is with the Engineering Research Center of the Ministry of Education for Intelligent Control System and Intelligent Equipment and Key Laboratory of Industrial Computer Control Engineering of Hebei Province, Yanshan University, Qinhuangdao 066004, China (e-mail: kma@ysu.edu.cn).}}

%
%
%\markboth{Journal of \LaTeX\ Class Files,~Vol.~14, No.~8, August~2015}%
%{Shell \MakeLowercase{\textit{et al.}}: Bare Demo of IEEEtran.cls for IEEE Journals}
%%%%%%%%%%%%%%%%%%%%%%%%%%%%%%%%%%%%%%%%%%%曹师姐作者模板
\maketitle

% As a general rule, do not put math, special symbols or citations
% in the abstract or keywords.
\begin{abstract}
This paper presents a carbon-energy coupling management framework for an industrial park, where the carbon flow model accompanying multi-energy flows is adopted to track and suppress carbon emissions on the user side. To deal with the quadratic constraint of gas flows, a bound tightening algorithm for constraints relaxation is adopted. The synergies among the carbon capture, energy storage, power-to-gas further consume renewable energy and reduce carbon emissions. 
Aiming at carbon emissions disparities and supply-demand imbalances, this paper proposes a carbon trading ladder reward and punishment mechanism and an energy trading and scheduling method based on Lyapunov optimization and matching game to maximize the long-term benefits of each industrial cluster without knowing the prior information of random variables. Case studies show that our proposed trading method can reduce overall costs and carbon emissions while relieving energy pressure, which is important for Environmental, Social and Governance (ESG).

\end{abstract}

% Note that keywords are not normally used for peerreview papers.

% For peer review papers, you can put extra information on the cover
% page as needed:
% \ifCLASSOPTIONpeerreview
% \begin{center} \bfseries EDICS Category: 3-BBND \end{center}
% \fi
%
% For peerreview papers, this IEEEtran command inserts a page break and
% creates the second title. It will be ignored for other modes.The industries in the park can be divided into three categories according to the characteristics of energy utilization: process industries, discrete manufacturing industries and emerging research and development industries \cite{Tang2018Morphological}. 
\IEEEpeerreviewmaketitle

\section{Introduction}

Nowadays, a large number of industrial clusters (ICs) have settled in the industrial park, which brings huge economic benefits, but forms a concentrated area with high energy consumption and emissions \cite{Wei2021Distribution}. To reduce emissions and save energy, the industrial park needs to be equipped with green distributed energy systems to improve the energy structure and supply efficiency.
Distributed multi-energy systems are built within ICs, which cooperates with gas companies and power grids to satisfy the diverse energy needs of users and achieve energy cascade utilization. The significant role of multi-energy coordination in improving the reliability, economy and cleanliness of energy systems has received extensive attention. % \cite{Zhu2022Energy} \cite{Li2021Optimal}\cite{Xu2019Distributed}Firstly, multi-energy coordination reduces the impact of the uncertainty of a single energy source and improves energy reliability \cite{Zhu2022Energy}; secondly, it realizes complementary substitution and cascade utilization of energy to improve economy \cite{Li2021Optimal}; finally, it compensates for the randomness of renewable energy and promotes full consumption \cite{Xu2019Distributed}.

Coal-fired power generation is gradually being replaced by natural gas and renewable power generation \cite{Liu2022Asynchronous}. As the increase of gas power generation, the correlation between electricity price and gas price increases. Gas prices affect power generation quotations, unit combinations and optimal scheduling \cite{Gil2016Electricity}, and electricity prices also affect natural gas production and transmission costs. Changes in electricity and natural gas prices lead to arbitrage behavior, which stimulates electricity and natural gas trading. The power-to-gas (P2G) technology and the increasing share of gas in electricity generation have strengthened the near real-time operational link between gas and electricity systems, which facilitates short-term trades in gas \cite{Pinson2017Towards}.

To increase overall efficiency, unlock synergy potential, and equitably distribute multi-energy generation resources, some countries are moving towards liberalization of energy markets \cite{Paepe2007Combined}. In Sweden, large-scale boilers and heat pumps are introduced into systems to use excess energy \cite{Averfalk2017Large}. A project is launched in Germany to tap the potential of heat pumps to increase the utilization of wind energy \cite{Papae2012Potential}. The deployment of boilers and heat pumps in Denmark over past decade and the decision to cut taxes in 2013 have created a favorable environment for the electrification of heating \cite{Nielsen2016Economic}. Although the coupled multi-energy market is immature in current practice, the marketing and operational challenges of integrated energy systems have received considerable attention.

Some studies have been conducted on multi-energy trading based on game or auction in industrial and residential parks to improve energy benefit and structure \cite{Zou2022Transactive}. In \cite{Chen2022Strategic}, an equilibrium model is used to deal with the noncooperative game problem of multi-energy producers, which helps to maximize the energy profit. In \cite{Xu2021Peer}, a multi-energy game-theoretic bargaining framework is proposed to coordinate multilateral resource and enhance the operational economy and resource utilization. In \cite{Liu2020Heat}, a multi-energy management framework based on Stackelberg game for heat and electricity trading in an industrial park is constructed to shift the peak load. In \cite{Zhong2018Auction}, auction mechanisms for multi-energy trading are proposed to integrate different energy and supply diverse energy for users. 
{These studies on multi-energy trading effectively achieve complementary sharing among energy sources, but lack consideration of carbon emission reduction among multi-energy ICs with different characteristics, which is indispensable for sustainable development.} %The few existing studies solve the multi-energy trading problem by matching game among ICs.%The carbon emission reduction of the system needs to be improved. This paper investigates the coordinated operation of the carbon capture and multi-energy devices, and proposes joint energy trading and scheduling method combining double auction and game. 

With the increasingly serious greenhouse effect caused by the burning of fossil fuels, the issue of emission reduction has attracted widespread attention. In fact, the establishment of a refined and reasonable carbon emission market is of great significance to promote energy marketization and emission reduction \cite{Zhang2019Carbon}. The International Financial Reporting Interpretations Committee describes emission rights as: the government grants participating entities the rights to emit a specific emission level \cite{Bradbury2007An}. Since carbon emission right can be purchased and sold to generate income, an active carbon emission market is formed. Chen et al. \cite{Chen2021Conjectural} provide a model of the interrelationships of the gas, electricity and carbon markets and use a relaxed method to deal with the secondary gas flow model. Huang et al. \cite{Huang2020Multienergy} introduce a carbon emission flow model to help multi-energy system planning to achieve low-carbon operation, and propose a generalized circuit theory to deal with the nonlinear model of gas flows. Wei et al. \cite{Wei2022Carbon} install the wind and photovoltaic (PV) power devices according to system planning to reduce the emissions, and use a linearized model to simplify the secondary function of gas flow between nodes to improve the computability of the planning. Wang et al. \cite{Wang2022Low} establish a low-carbon operation model in a multi-energy system and utilize the carbon capture power plant (CCPP) and P2G technology to greatly reduce the emissions, but ignore the nonlinearity of gas flows. {These related studies have realized the economic operation of low-carbon integrated energy systems, but seldom consider the comprehensive impact of multi-energy chain on the environment, the complexity of multi-energy coordination and the nonlinearity of gas flows.} %This paper analyzes the carbon emissions of each energy chain to achieve the economic operation of a low-carbon integrated energy system under carbon and multi-energy framework, and proposes a bound tightening algorithm to deal with the nonlinear of gas flows. proposes a two-level planning model for multi-energy systems, and 

Considering that end users are potential influencers of carbon emissions, so emissions from energy generation should be considered from the demand-side perspective. Kang et al. \cite{Kang2015Carbon} introduce an emission flow model to clarify the users' emission responsibility. Cheng et al. \cite{Cheng2019Planning} adopt the carbon emission flow model to allocate emissions from a consumption perspective. Shen et al. \cite{Shen2020Low} apply the emission model to calculate demand-side emissions to evaluate the effect of the low-carbon transformation. Wang et al. \cite{Wang2020Carbon} adopt a management model to achieve emission reduction by encouraging users to participate in electricity and carbon trading.
%Ref. \cite{Kang2015Carbon} introduces a carbon emission flow model in the power network to clarify the carbon emission responsibility of users. Ref. \cite{Cheng2019Planning} adopts the carbon flow model to allocate carbon emissions from the perspective of consumption and coordinate inter-regional and intra-regional planning. Ref. \cite{Wang2020Low} uses a carbon emission flow model to track carbon emissions, and adopts a two-stage low-carbon operation planning model to focus on carbon emissions reduction. Ref. \cite{Wang2020Carbon} adopts a two-stage scheduling model to achieve carbon emission reduction by encouraging consumers to participate in electricity and carbon emissions trading.
{Although the carbon emission flow models achieve a more accurate calculation of carbon emissions, the above research lacks consideration of the spatiotemporal coupling characteristics of carbon and multi-energy, and the interaction of multi-energy and carbon trading. Carbon and energy coupling means that different sources of energy produce different carbon emissions. For example, renewable energy generation produces no carbon emissions, while coal-fired generation produces high emissions. When ICs trade energy, the carbon emissions in energy is also transferred. This paper takes this into account in carbon and energy trading. } %This paper builds a carbon-gas-electricity coordinated trading network model and achieves hierarchical decoupling of transactions at different timescales.

%An industrial park is often composed of multiple ICs, which can be divided into three categories according to the characteristics of energy utilization: process industries, discrete manufacturing industries and emerging research and development industries. However, many studies only carry out energy planning for a single IC, ignoring the coordination and complementarity of energy consumption among various ICs, which results in energy waste and high carbon emissions. Therefore, this paper explores the potential for energy synergy among multiple typical ICs and studies the joint benefits of demand-side in coupling carbon, gas and electricity trading markets. 

%At present, one of the gaps to trading markets is the difference in the timescale of the industrial energy system. For instance, the generation and transmission of heat and gas need to take a few minutes even hours to achieve energy balance, while the electricity generation and transmission is instantaneous. Thus, the carbon, gas and electricity trading is constructed as a three-timescale optimization problem to shift carbon emission and improve energy efficiency in this paper. The timescales of carbon, gas and electricity trading are a day, one hour and 15 minutes, respectively. Although the timescales of the three markets are different, there are multi-energy couplings and implicit carbon costs in multi-energy markets; while multi-energy scheduling also affects the benefit or cost of carbon trading.

{In addition, long-term stochastic optimization problems need to be solved in energy trading. %Nowadays, most existing trading algorithms rarely consider long-term storage constraint balance when dealing with stochastic optimization problems.
Different from traditional methods like dynamic programming, which requires prior information about the  random processes in the system, Lyapunov optimization can give simple online solution to minimize cost and queue fluctuation. Some Lyapunov optimization-based algorithms achieve efficient energy management by transforming the long-term stochastic problem into deterministic subproblems for each time slot }\cite{Li2020A}. {However, it remains difficult to make optimal online scheduling decisions in carbon and energy trading.
 To achieve effective carbon and energy management among ICs, a multi-energy trading method and a ladder carbon trading mechanism are jointly considered to make optimal scheduling decisions. }
Compared to existing research, our contributions are summarized.
\begin{enumerate}
\item
%A coupled carbon and energy management framework including a carbon capture power plant (CCPP), P2G devices, CHP units, boilers, energy storage and photovoltaic (PV) panels for ICs is proposed. The CCPP, P2G devices and energy storage synergize to further consume renewable energy, reduce carbon emissions, and achieve effective energy management. The carbon emissions of ICs are tracked and calculated based on the emissions accompanying multi-energy flows. 
A coupled carbon and multi-energy management framework including a CCPP, P2G devices and energy storage for ICs is proposed. The synergies among the CCPP, P2G devices and energy storage further consume renewable energy, reduce carbon emissions, and improve overall benefits. A carbon flow model accompanying multi-energy flows is adopted to track and suppress carbon emissions on the user side. 
%The carbon emissions of ICs are tracked and calculated based on the emissions accompanying multi-energy flows. 

\item 
%To make full use of the difference in carbon emission and the alternative complementarity of multiple energy sources, a carbon and energy coordinated trading model is constructed to improve the environmental and economic benefits of the ICs, where the reward and punishment mechanisms of carbon trading is adopted to improve the willingness to trade. The trading is hierarchical decoupled on multiple timescales according to the characteristic of carbon emission and multi-energy. on multiple timescales
To make full use of the differences in carbon emissions and the alternative complementarity of energy, a carbon and energy coordinated trading model is constructed to improve the benefits of the ICs, where the carbon trading ladder reward and punishment mechanism is adopted to improve the willingness to trade. A boundary tightening algorithm is adopted to increase the tightness of the quadratic constraint relaxation of gas flows. 

\item 
%A bound tightening algorithm is proposed to improve the tightness of the relaxation for the quadratic constraint of gas flows. A multi-energy trading and scheduling method combining game and double auction is proposed, where the game among sellers determines the optimal policy of energy to sell, and the double auction determines the deal prices of sellers and buyers. The time delay of heat and gas is considered to deal with the heterogeneity of multi-energy. 
%To improve the imbalance of supply and demand and utilize the spatiotemporal availability and cost diversity of energy, a joint trading and scheduling algorithm based on Lyapunov optimization and matching game is proposed, which minimizes the long-term cost of each IC without knowing the prior information of random variables. The matching game provides low-complexity distributed solutions to the energy matching problem. 
To alleviate the supply and demand imbalance of individual ICs, a joint trading and scheduling method based on Lyapunov optimization and matching game is proposed, which minimizes the cost of each IC without knowing the prior information of random variables. The matching game provides low-complexity distributed solutions to the energy matching problem between buyers and sellers. 
% . %Moreover, Simulation results based on real data show that the coordination of carbon capture, carbon and energy trading reduces energy costs and carbon emissions, while alleviating energy pressure. 
\end{enumerate}

The rest of the work is as follows. The multi-energy devices, carbon emission flow model, energy and carbon trading model among ICs are introduced in Section \uppercase\expandafter{\romannumeral2}. The constraints relaxation of gas flows, joint trading and scheduling method and performance analysis are proposed in Section \uppercase\expandafter{\romannumeral3}. %, and its distributed fast realization is presented in Section \uppercase\expandafter{\romannumeral4}. 
%The performance analysis of the proposed distributed stochastic gradient algorithm is provided in Section \uppercase\expandafter{\romannumeral4}. 
Section \uppercase\expandafter{\romannumeral4} gives the simulation results. Finally, Section \uppercase\expandafter{\romannumeral5} concludes the paper and presents further research.

\section{System Model}
\begin{figure}
  \centering
  \includegraphics[width=.7\hsize]{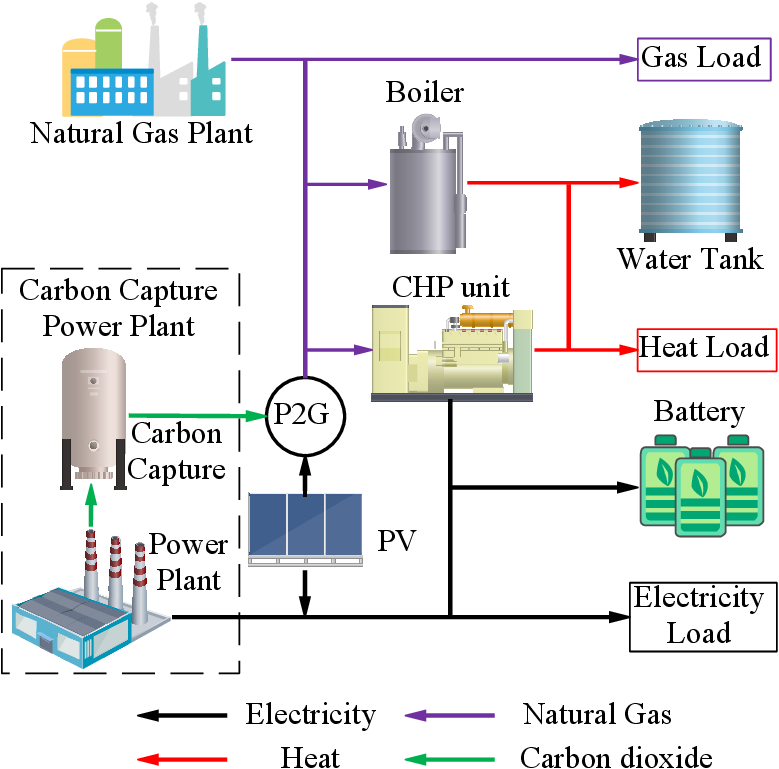}
  \caption{Multi-energy supply of ICs}
  \label{fig1}
\end{figure}

%\subsection{Industrial Park}
In this paper, a system including an industrial park, a CCPP and a natural gas plant is considered, as shown in Fig.~\ref{fig1}. The CCPP consists of a carbon capture device and a power plant, and the amount of carbon capture can be adjusted by changing the power of the carbon capture device. The industrial park consists of ICs and multi-energy devices, which includes CHP units, PV panels, batteries, water tanks, boilers and P2G devices. The industrial park can obtain heat and electricity from CHP units, and store excess energy in water tanks and batteries for the future energy demands. 
% and renewable energy generated by PV panelsThe park has $\boldsymbol{K}=\{1, 2, ..., K\}$ MEGPs. The energy equipments in the next subsection are modeled for MEGP $k$, $k\in \boldsymbol{K}$. %A list of notations is shown in Table I, where one time slot is one hour to coordinate with the simulation.\cite{Yazdani2018Strategic}

%\mbox{}
\nomenclature{$t$}{Time slot, $t\in \mathbb{N}$, where $\mathbb{N}$ is the set of natural numbers.}
\nomenclature{$i$}{IC, $i\in \{1,2,\cdots,N\}$.}
\nomenclature{$B_{i}(t)$}{Electricity of battery $i$.}
\nomenclature{$W_{i}(t)$}{Thermal energy of water tank $i$.}
\nomenclature{$C_{i}^{\rm{e}}(t),D_{i}^{\rm{e}}(t)$}{Charge and discharge of battery $i$.}
\nomenclature{$C_{i}^{\rm{h}}(t),D_{i}^{\rm{h}}(t)$}{Thermal energy charged and discharged by water tank $i$.}
%$D_{i}^{\rm{e}}(t)$ & The electricity discharged from battery $k$ at time t.\\
%$D_{i}^{\rm{h}}(t)$ & The thermal energy discharged from hot water tank $k$.\\
%$\eta_{cke}, \eta_{ckh}$ & The charging efficiencies of battery $k$ and water tank $k$.\\
%$\eta_{dke}, \eta_{dkh}$ & The discharging efficiencies of battery $k$ and water tank $k$.\\
\nomenclature{$E_{i}^{\rm{CHP}}(t),H_{i}^{\rm{CHP}}(t),G_{i}^{\rm{CHP}}(t)$}{Electricity generation, heat generation and gas consumption of CHP unit $i$.}
%\nomenclature{$H_{i}^{\rm{CHP}}(t)$}{Heat generation of CHP unit $i$.}
%\nomenclature{$G_{i}^{\rm{CHP}}(t)$}{Gas consumption of CHP unit $i$.}
%$\eta_{kpg}, \eta_{khg}$ & The conversion efficiencies of CHP unit $k$ from gas to electricity and heat.\\
\nomenclature{$H_{i}^{\rm{b}}(t),G_{i}^{\rm{b}}(t)$}{Heat generation and gas consumption of boiler $i$.}
%$G_{i}^{\rm{b}}(t)$ & The gas consumption of boiler $i$.\\
\nomenclature{$E_{i}^{\rm{r}}(t)$}{Electricity generated by PV panel $i$.}
%$G_{i}^{\rm{p2g}}(t)$ & The gas generated by P2G $i$.\\
\nomenclature{$E_{i}^{\rm{p2g}}(t),G_{i}^{\rm{p2g}}(t)$}{Electricity consumption and gas generation of P2G $i$.}
%$I_{n}^{\rm{eN}},I_{n}^{\rm{gN}}$ & The carbon intensities of node $n$ in the electricity and gas network.\\
%$E_{m}$ & The electricity flow of line $m$.\\
%$I_{m}^{\rm{e}}$ & The carbon intensity of electricity line $m$.\\
%$E_{n}^{\rm{G}}$ & The electricity output of generator $n$.\\
%$I_{n}^{\rm{G}}$ & The carbon intensity of generator $n$.\\
%$R_{m}^{\rm{e}}$ & The carbon emission of electricity line $m$.\\
%$R_{n}^{\rm{G}}$ & The carbon emission of generator $n$.\\
%$R_{c}$ & The carbon emission of compressor $c$.\\
%$E_{c}^{\rm{o}}$ & The electricity consumption of compressor $c$ at time t.\\
%$R_{p}^{\rm{g}},I_{p}^{\rm{g}},f_{p}^{\rm{g}}$ & The carbon emission, carbon intensity, gas flow of gas pipeline $p$ at time t.\\
\nomenclature{$f_{nm}$}{Gas flow from node $n$ to node $m$.}
\nomenclature{$C_{nm}$}{Weymouth constant for pipeline $nm$.}
\nomenclature{$\pi_{n},\pi_{m}$}{Gas pressures at node $n$ and node $m$.}
%$R^{\rm{gl}},I^{\rm{gl}},f^{\rm{gl}}$ & The carbon emission, carbon intensity, gas flow of gas load at time t.\\
\nomenclature{$Q_{i}^{\rm{a}}(t),Q_{i}^{\rm{d}}(t)$}{Actual and free carbon emissions of IC $i$.}
%$I_{i}^{\rm{el}}(t),I_{i}^{\rm{gl}}(t),I_{i}^{\rm{hl}}(t)$ & The carbon intensity of electricity, gas and thermal load of IC $i$.\\
\nomenclature{$E_{i}^{\rm{d}}(t),G_{i}^{\rm{d}}(t),H_{i}^{\rm{d}}(t)$}{Total electricity, gas and thermal loads of IC $i$.}
%$\eta_{kbg}$ & The conversion efficiency of boiler $k$ from gas to heat.\\
\nomenclature{$E_i(t),p^{\rm{e}}(t)$}{Electricity and price purchased from the CCPP.}
\nomenclature{$C_{i}^{\rm{c}}(t)$}{Carbon trading cost of IC $i$.}
\nomenclature{$C_{i}^{\rm{X}}(t),R_{i}^{\rm{S}}(t)$}{Cost and revenue of IC $i$ for multi-energy trading among ICs.}
\nomenclature{$p_{ij}^{\rm{e}}(t),p_{ij}^{\rm{g}}(t)$}{Electricity and gas prices that buyer $i$ purchases from seller $j$.}
\nomenclature{$s_{ij}^{\rm{e}}(t),s_{ij}^{\rm{g}}(t)$}{Electricity and gas prices that seller $i$ sells to buyer $j$.}
\nomenclature{$G_{i}(t),p^{\rm{g}}(t)$}{Gas and price purchased from the natural gas plant.}
\nomenclature{$E_{i}^{\rm{o}}(t),p^{\rm{o}}(t)$}{Electricity and price sold back to the CCPP.}
\nomenclature{$X_{i}^{\rm{eb}}(t),X_{i}^{\rm{es}}(t)$}{Electricity purchased and sold by IC $i$ through multi-energy trading among ICs.}
\nomenclature{$X_{i}^{\rm{gb}}(t),X_{i}^{\rm{gs}}(t)$}{Gas purchased and sold by IC $i$ through multi-energy trading among ICs.}
\nomenclature{$C_{i}(t)$}{Total cost of IC $i$.}
%\nomenclature{$p^{\rm{e}}(t)$}{Electricity price of the CCPP.}
%\nomenclature{$p^{\rm{g}}(t)$}{Gas price of the gas plant.}
%\nomenclature{$p^{\rm{o}}(t)$}{Electricity price sold to the CCPP.}
\nomenclature{$C_i^{\rm{opt}}(t),C_{ir}^{\rm{opt}}(t)$}{Optimal costs for the original problem and the relaxed problem.}
\nomenclature{$F_{i}(t),Z_{i}(t)$}{Battery and water tank virtual queues for IC $i$.}
\nomenclature{$\theta_{i}(t),\epsilon_{i}(t)$}{Perturbation terms to ensure battery and water tank constraints.}
\nomenclature{$V_{i}$}{Tradeoff term between energy cost and queue stability.}
\nomenclature{$\pi_{nm}^{\rm{d}}(t),\pi_{nm}^{\rm{s}}(t)$}{Difference and sum of pressures between nodes $n$ and $m$.}
%$W_{i}(t)$ & Thermal energy of hot water tank $i$.\\
%\nomenclature{\{B_{i}(t)\}}{Electricity of battery $i$.}\\
\printnomenclature

\subsection{Multi-energy Devices}
The ICs are equipped with batteries, water tanks, CHP units, boilers, PV panels and P2G devices to supply energy for users according to their energy consumption characteristic. {For IC $i$, $\forall i \in \{1,2,\cdots,N\}$, where $N$ denotes the quantity of ICs in the industrial park, the battery of IC $i$ is denoted as battery $i$, and the same applies to other devices below. The electricity of the battery $i$ is $B_{i}(t)$ and the equivalent thermal energy of the hot water tank $i$ is $W_{i}(t)$, where the time slot $\forall t \in \mathbb{N}$ with $\mathbb{N}$ being the set of natural numbers, and one time slot is a hour.} The dynamic models of battery $i$ and water tank $i$ are 
\begin{equation}
B_{i}(t+1)=B_{i}(t)+C_{i}^{\rm{e}}(t)-D_{i}^{\rm{e}}(t), \forall i,t
\label{A1}
\end{equation}
\begin{equation}
W_{i}(t+1)=W_{i}(t)+C_{i}^{\rm{h}}(t)-D_{i}^{\rm{h}}(t), \forall i,t
\label{A3}
\end{equation}
\begin{equation}
B_{i}^{\rm{min}}\leq B_{i}(t) \leq B_{i}^{\rm{max}}, W_{i}^{\rm{min}}\leq W_{i}(t) \leq W_{i}^{\rm{max}}, \forall i,t
\label{Wm}
\end{equation}
\begin{equation}
0\leq C_{i}^{\rm{e}}(t) \leq C_{i}^{\rm{e,max}}, 0\leq D_{i}^{\rm{e}}(t) \leq {D_{i}^{\rm{e},\max}}, \forall i,t
\label{Cem}
\end{equation}
\begin{equation}
0\leq C_{i}^{\rm{h}}(t) \leq C_{i}^{\rm{h,max}}, 0\leq D_{i}^{\rm{h}}(t) \leq D_{i}^{\rm{h,max}}, \forall i,t
\label{Chm}
\end{equation}
where $C_{i}^{\rm{e}}(t)$ and $D_{i}^{\rm{e}}(t)$ denote the charging and discharging amount of battery $i$. $C_{i}^{\rm{h}}(t)$ and $D_{i}^{\rm{h}}(t)$ are the thermal energy charging and discharging amount of hot water tank $i$. %where $B_{k}(t)$, $\eta_{cke}$, $C_{ke}(t)$, $\eta_{dke}$ and $D_{ke}(t)$ denote the stored electricity, charging efficiency, charging rate, discharging efficiency and discharging rate of the battery. $W_{k}(t)$, $\eta_{ckh}$, $C_{kh}(t)$, $\eta_{dkh}$ and $D_{kh}(t)$ denote the stored thermal energy, charging efficiency, charging rate, discharging efficiency and discharging rate of the water tank. 
%\subsubsection{CHP}
%A CHP unit generates heat and electricity simultaneously, which can be denoted by
%\subsection{Feasible Region for CHP}

\begin{figure}
\includegraphics[width=\hsize]{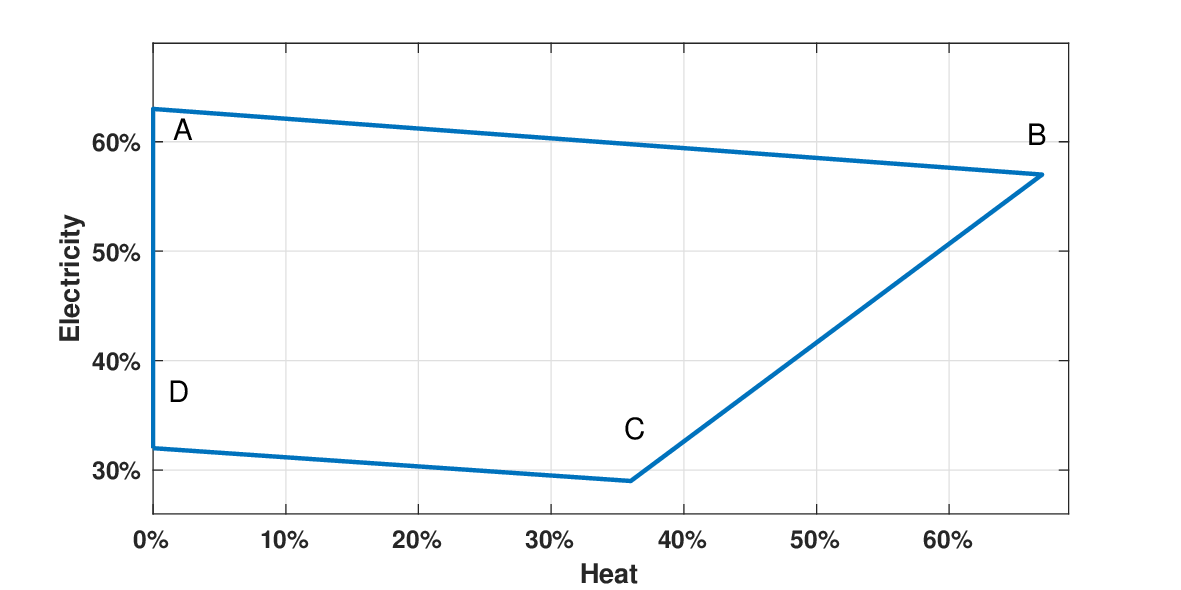}
\caption{feasible region for CHP}
\end{figure}
{The CHP unit, as a co-generation unit, can generate both heat and electricity by consuming natural gas $G_{i}^{\rm{CHP}}(t)$. However, the generation of heat $H_{i}^{\rm{CHP}}(t)$ and electricity $E_{i}^{\rm{CHP}}(t)$ cannot be separately considered. The co-generation is coupled and interdependent.  The generation constraints of the CHP unit are operated in a feasible region which is depicted in Fig. 2} \cite{Chen2015Increasing}. {The feasible operating region is bounded by curve ABCD. A, B, C and D represent four marginal points of the feasible region for the CHP unit. In order to describe the feasible region of CHP unit $i$, we have the following inequality constraints }
\begin{equation}
\begin{aligned}
&E_{i}^{\rm{CHP}}(t)-E_{i}^{\rm{CHP,A}}-\frac{E_{i}^{\rm{CHP},\rm{A}}-E_{i}^{\rm{CHP,B}}}{H_{i}^{\rm{CHP,A}}-H_{i}^{\rm{CHP,B}}}(H_{i}^{\rm{CHP}}(t)\\&-H_{i}^{\rm{CHP,A}}) \leq 0, \forall i,t
\end{aligned}
\label{eq1}
\end{equation}
\begin{equation}
\begin{aligned}
&E_{i}^{\rm{CHP}}(t)-E_{i}^{\rm{CHP,B}}-\frac{E_{i}^{\rm{CHP,B}}-E_{i}^{\rm{CHP,C}}}{H_{i}^{\rm{CHP,B}}-H_{i}^{\rm{CHP,C}}}(H_{i}^{\rm{CHP}}(t)\\&-H_{i}^{\rm{CHP,B}}) \geq 0, \forall i,t
\end{aligned}
\label{eq2}
\end{equation}
\begin{equation}
\begin{aligned}
&E_{i}^{\rm{CHP}}(t)-E_{i}^{\rm{CHP,C}}-\frac{E_{i}^{\rm{CHP,C}}-E_{i}^{\rm{CHP,D}}}{H_{i}^{\rm{CHP,C}}-H_{i}^{\rm{CHP,D}}}(H_{i}^{\rm{CHP}}(t)\\&-H_{i}^{\rm{CHP,C}}) \geq 0, \forall i,t
\end{aligned}
\label{eq3}
\end{equation}
\begin{equation}
0 \leq H_{i}^{\rm{CHP}}(t) \leq H_{i}^{\rm{CHP,B}}, \forall i,t
\end{equation}
\begin{equation}
0 \leq E_{i}^{\rm{CHP}}(t) \leq E_{i}^{\rm{CHP,A}}, \forall i,t
\label{eq5}
\end{equation}

Eq. (\ref{eq1}) denotes the area under the curve AB. Eq. (\ref{eq2}) and Eq. (\ref{eq3}) denote the area above the curve BC and CD respectively.

%The amount of natural gas consumed, heat and electricity generated by CHP unit $i$ are $G_{iCHP}(t)$, $H_{iCHP}(t)$ and $E_{iCHP}(t)$. The amount of gas consumed and heat generated by boiler $i$ are $G_{ib}(t)$ and $H_{ib}(t)$. The CHP unit $i$ and boiler $i$ models are denoted as
%\begin{equation}
%E_{iCHP}(t)=\eta_{ipg}G_{iCHP}(t), H_{iCHP}(t)=\eta_{ihg}G_{iCHP}(t)
%\label{CHP1}
%\end{equation}
%\begin{equation}
%0\leq E_{iCHP}(t) \leq E_{iCHP,max}, 0\leq H_{iCHP}(t) \leq H_{iCHP,max}
%\label{CHP2}
%\end{equation}
%where $E_{kCHP}(t)$, $\eta_{kpg}$, $G_{kCHP}(t)$, $H_{kCHP}(t)$ and $\eta_{khg}$ are the electricity generation, electricity generation efficiency, natural gas consumption, heat generation and heat generation efficiency of the CHP unit at time slot $t$, respectively. 
%\subsubsection{Boiler}
The boiler $i$ generates heat $H_{i}^{\rm{b}}(t)$ by consuming natural gas $G_{i}^{\rm{b}}(t)$, which is 
%The amount of gas consumed and heat generated by boiler $i$ are $G_{ib}(t)$ and $H_{ib}(t)$
\begin{equation}
H_{i}^{\rm{b}}(t)=\eta_{i}^{\rm{bg}}G_{i}^{\rm{b}}(t), 0\leq H_{i}^{\rm{b}}(t) \leq H_{i}^{\rm{b,max}}, \forall i,t
\label{bo1}
\end{equation}
%where $\eta_{ipg}$ and $\eta_{ihg}$ denote the efficiencies of CHP unit $i$ to generate electricity and heat. 
where $\eta_{ibg}$ denotes the efficiency of boiler $i$ to generate heat. %\begin{equation}
%0\leq H_{ib}(t) \leq H_{ib,max}
%\label{bo2}
%\end{equation}
%\begin{equation}
%E_{ic}(t) = \eta_{ice}O_{ic}
%\label{ic1}
%\end{equation}

The amount of electricity generated by PV panel $i$ is $E_{i}^{\rm{r}}(t)$. % and $E_{i}(t)$. %The amount of coal consumed and electricity generated by CCPP $i$ are denoted as $O_{ic}(t)$ and $E_{ic}(t)$. 
The amount of electricity consumed and gas generated by P2G device $i$ are $E_{i}^{\rm{p2g}}(t)$ and $G_{i}^{\rm{p2g}}(t)$.
%\subsubsection{Energy Storage Systems}
 The models of PV panel $i$ and P2G $i$ are denoted as 
 \begin{equation}
0\leq E_{i}^{\rm{r}}(t) \leq E_{i}^{\rm{r,max}}, \forall i,t
\label{ic2}
\end{equation}
%\begin{equation}
%0\leq E_{ic}(t) \leq E_{ic,max}
%\label{ic2}
%\end{equation}
\begin{equation}
G_{i}^{\rm{p2g}}(t) = \eta_{i}^{\rm{p2g}}E_{i}^{\rm{p2g}}(t), 0\leq G_{i}^{\rm{p2g}}(t) \leq G_{i}^{\rm{p2g,max}}, \forall i,t
\label{p2g1}
\end{equation}
%\begin{equation}
%0\leq G_{ip2g}(t) \leq G_{ip2g,max}
%\label{p2g2}
%\end{equation}
where $\eta_{i}^{\rm{p2g}}$ is the efficiency of P2G $i$. %$\eta_{ice}$ is the conversion efficiency of CCPP $i$ from coal to electricity. 
\subsection{Carbon Emission Flow Model}
{In order to obtain the emissions corresponding to energy consumption on the consumer side, and track the carbon footprints of different energy forms in the multi-energy system of the industrial park, a carbon flow model accompanying multi-energy flows is introduced to calculate the emissions of ICs}\cite{Cheng2019Modeling}. 
\subsubsection{Carbon Emission Accompanying Electricity Flow}
\begin{figure}
  \centering
  \includegraphics[width=.8\hsize]{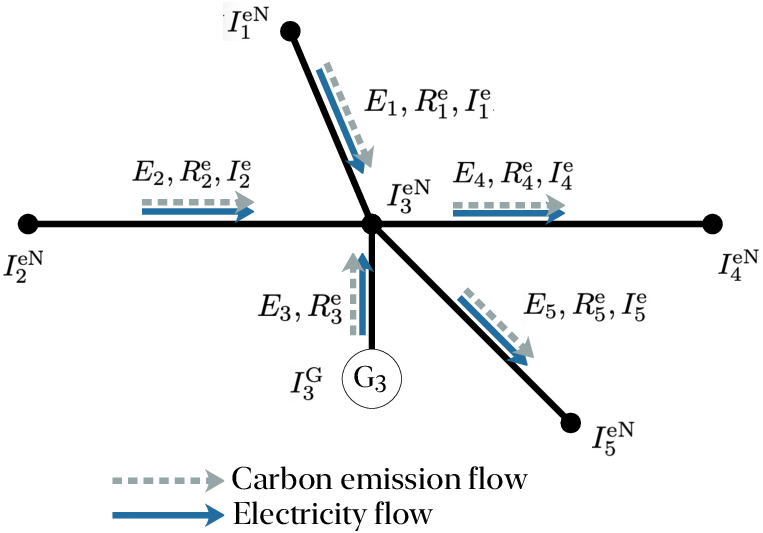}
  \caption{Carbon emission accompanying the electricity flow}
  \label{f2}
\end{figure}

{According to the basic principles of }\cite{Cheng2019Modeling}, the carbon emission intensity of a node is determined by the emissions flowing into the node, which is equal to the carbon emission intensity of the outflow line of the node. Taking node 3 in electricity network as an example in Fig.~\ref{f2}, the carbon emission intensity $I_{3}^{\rm{eN}}$ of node 3 equals the carbon intensity of lines 4 and 5, which is denoted by
\begin{equation}
\begin{aligned}
I_{3}^{\rm{eN}}&=\frac{E_1I_{1}^{\rm{e}}+E_2I_{2}^{\rm{e}}+E_{3}^{\rm{G}}I_{3}^{\rm{G}}}{E_1+E_2+E_{3}^{\rm{G}}}=\frac{R_{1}^{\rm{e}}+R_{2}^{\rm{e}}+R_{3}^{\rm{G}}}{E_1+E_2+E_{3}^{\rm{G}}}=I_{4}^{\rm{e}}=I_{5}^{\rm{e}}
\label{cef}
\end{aligned}
\end{equation}
where $E_1$, $E_2$ and $E_{3}^{\rm{G}}$ denote the electricity flows of lines 1, 2 and generator 3. $I_{1}^{\rm{e}}$, $I_{2}^{\rm{e}}$ and $I_{3}^{\rm{G}}$ denote the carbon emission intensities of lines 1, 2 and generator 3. $R_{1}^{\rm{e}}$, $R_{2}^{\rm{e}}$  and $R_{3}^{\rm{G}}$ denote the carbon emissions of lines 1, 2 and generator 3. 

According to the carbon intensity at node 3, the carbon intensity at node $n$ is deduced by
\begin{equation}
I_{n}^{\rm{eN}}=\frac{\sum_{m\in M}E_mI_{m}^{\rm{e}}+E_{n}^{\rm{G}}I_{n}^{\rm{G}}}{\sum_{m\in M}E_m+E_{n}^{\rm{G}}}=\frac{\sum_{m\in M}R_{m}^{\rm{e}}+R_{n}^{\rm{G}}}{\sum_{m\in M}E_m+E_{n}^{\rm{G}}}, \forall n
\label{cefn}
\end{equation}
where $M$ is the line set of emissions flowing into node $n$.

\subsubsection{Carbon Emission Accompanying Gas Flow}
{In the natural gas pipeline, compressors are equipped to compensate the pressure losses caused by resistance to ensure the reliable gas delivery. }The carbon emission $R_{c}$ of compressor $c$ is denoted as
\begin{equation}
R_{c}=I_{c}^{\rm{eN}}E_{c}^{\rm{o}}, \forall c
\label{cec}
\end{equation}
where $I_{c}^{\rm{eN}}$ and $E_{c}^{\rm{o}}$ denote the nodal carbon intensity and electricity consumption of compressor $c$. The emissions $R_{p}^{\rm{g}}$ of gas pipeline $p$ are the sum of the emissions from the compressor and the emissions accompanying the gas flow, which is denoted as
\begin{equation}
R_{p}^{\rm{g}}=I_{p}^{\rm{g}}f_{p}^{\rm{g}}B+R_{c}, \forall p
\label{cegp}
\end{equation}
where $I_{p}^{\rm{g}}$ denotes the carbon intensity of gas pipeline $p$; $f_{p}^{\rm{g}}$ denote the gas flow of pipeline $p$; $B$ is the calorific value of gas, which is 10 kWh/m$^3$. The nodal carbon intensity $I_{n}^{\rm{gN}}$ of gas, which is similar as the one of electricity, is denoted as
\begin{equation}
I_{n}^{\rm{gN}}=\frac{\sum_pR_{p}^{\rm{g}}+f_wI^{\rm{gc}}B}{(\sum_pf_{p}^{\rm{g}}+f_w)B}, \forall n,w
\label{ncig}
\end{equation}
where $I^{\rm{gc}}$ denotes the carbon emission per unit of gas combustion, which is 0.2 kgCO$_2$/kWh. $f_w$ denotes the gas flow of gas source $w$ in node $n$. 
{The gas $f_{nm}$ flowing from node $n$ to node $m$ in the pipeline $\forall nm \in \Omega_p$ with $\Omega_p$ being the set of gas pipelines is denoted as}
\begin{equation}
 f^2_{nm}=C^2_{nm}(\pi^2_{n}-\pi^2_{m}), \forall nm
 \label{weco}
\end{equation}
\begin{equation}
\pi^{\min}\leq \pi \leq \pi^{\max}
\label{ipa}
\end{equation}
\begin{equation}
\rho^{\min}\leq \frac{\pi_n}{\pi_m}\leq \rho^{\max}, \forall nm
\end{equation}
where $C_{nm}$ denote the Weymouth constant for pipeline $nm$. $\pi_n$ and $\pi_m$ are the gas pressures at nodes $n$ and $m$. %C_{nm}=50kcf/psig=200m^3/KPa

The carbon intensity $I_{n}^{\rm{hN}}$ of heat load at node $n$ is similar to the carbon intensity of electricity load at node $n$.

\subsection{Carbon Trading}
Each IC $i$ is allocated free carbon emissions $Q_{i}^{\rm{d}}(t)$ based on industry benchmarks. 
%According to the emission factors for different types of energy consumption, the allocated carbon emissions of IC $i$ are 
%\begin{equation}
%Q_{id}(t)=a_{ie}E_{it}(t)+a_{ig}G_{it}(t)+a_{ih}H_{it}(t)
%\label{fcea}
%\end{equation}
%where $a_{ie}$, $a_{ig}$ and $a_{ih}$ are the allocated carbon emissions per unit of electricity, gas and heat. $E_{it}(t)$, $G_{it}(t)$ and $H_{it}(t)$ are the total electricity, gas and heat demand of IC $i$ at time $t$.

The actual emissions are related to the energy load and its carbon intensity, where the CCPP can capture CO$_2$ to reduce the actual emission of ICs. The actual emissions of IC $i$ are 
\begin{equation}
\begin{aligned}
%\mathcolorbox{yellow}
Q_{i}^{\rm{a}}(t)&=\sum_{n \in L_i^{\rm{e}}} I_{n}^{\rm{eN}}(t)E_{n}^{\rm{d}}(t)+\sum_{n \in L_i^{\rm{g}}} I_{n}^{\rm{gN}}(t)G_{n}^{\rm{d}}(t)B\\&+\sum_{n \in L_i^{\rm{h}}} I_{n}^{\rm{hN}}(t)H_{n}^{\rm{d}}(t)-\eta^{\rm{cc}}b^{\rm{cc}}E_{i}^{\rm{cc}}(t),      \forall n,t
\label{acei}
\end{aligned}
\end{equation}
{where $L_{i}^{\rm{e}}$, $L_{i}^{\rm{g}}$ and $L_{i}^{\rm{h}}$ are the electricity, gas and thermal load node sets of IC $i$. $E_{n}^{\rm{d}}(t)$, $G_{n}^{\rm{d}}(t)$ and $H_{n}^{\rm{d}}(t)$ are  electricity, gas and heat loads of node $n$. $\eta^{\rm{cc}}$, $b^{\rm{cc}}$ and $E_{i}^{\rm{cc}}(t)$ denote the carbon capture efficiency, unit carbon emissions and electricity generation of the CCPP for IC $i$.}  %$b_{cc}$ denotes the unit carbon emissions of the CCPP. $E_{i}$ denotes the electricity generation of the CCPP. 

%\\&+Q_{iCHP}(t)+Q_{ib}(t)+Q_{iP2G}(t)+Q_{icc}(t)$Q_{iCHP}(t)$, $Q_{ib}(t)$, $Q_{iP2G}(t)$, $Q_{icc}(t)$ denote the carbon emission of energy devices owing to the +I_{ith}(t)H_{it}(t)

To encourage ICs to reduce emissions and use green energy, the ladder reward and punishment carbon trading mechanism is introduced. The more the actual carbon emissions exceed the allocated carbon emissions, the faster trading costs increase. %The more the allocated carbon emissions exceed the actual carbon emissions, the faster the trading benefits increase. 
%Therefore, t
The carbon trading cost of IC $i$, $\forall i$, can be denoted in (\ref{cct}), where the time slot $t$ is removed for simplicity. %每天都有实际碳排放和分配的碳排放（由消耗的能量和相应的碳排放系数决定），因此每天都有对应的碳交易成本
%\begin{small}
\begin{equation}
\begin{aligned}
C_{i}^{\rm{c}}=\left\{\begin{array}{cc}
-p^{\rm{c}}(1+k\beta)(Q_{i}^{\rm{d}}-Q_{i}^{\rm{a}}-kl)-(k+\frac{(k-1)k\beta}{2})p^{\rm{c}}l,\\ \mbox{if}\  Q_{i}^{\rm{d}}-(k+1)l<Q_{i}^{\rm{a}}\leq Q_{i}^{\rm{d}}-kl \\
-p^{\rm{c}}(1+\beta)(Q_{i}^{\rm{d}}-Q_{i}^{\rm{a}}-l)-p_cl, \\\mbox{if}\  Q_{i}^{\rm{d}}-2l<Q_{i}^{\rm{a}}\leq Q_{i}^{\rm{d}}-l \\
-p^{\rm{c}}(Q_{i}^{\rm{d}}-Q_{i}^{\rm{a}}), \mbox{if}\  Q_{i}^{\rm{d}}-l<Q_{i}^{\rm{a}}\leq Q_{i}^{\rm{d}} \\
p^{\rm{c}}(Q_{i}^{\rm{a}}-Q_{i}^{\rm{d}}), \mbox{if}\  Q_{i}^{\rm{d}}<Q_{i}^{\rm{a}}\leq Q_{i}^{\rm{d}}+l \\
p^{\rm{c}}(1+\alpha)(Q_{i}^{\rm{a}}-Q_{i}^{\rm{d}}-l)+p_cl, \\\mbox{if}\  Q_{i}^{\rm{d}}+l<Q_{i}^{\rm{a}}\leq Q_{i}^{\rm{d}}+2l \\
p^{\rm{c}}(1+k\alpha)(Q_{i}^{\rm{a}}-Q_{i}^{\rm{d}}-kl)+(k+\frac{(k-1)k\alpha}{2})p^{\rm{c}}l,\\ \mbox{if}\  Q_{i}^{\rm{d}}+kl<Q_{i}^{\rm{a}}\leq Q_{i}^{\rm{d}}+(k+1)l \\
\end{array}\right.
\label{cct}
\end{aligned}
\end{equation}
%\end{small}
where $p^{\rm{c}}$ denotes the unit price of carbon emission allowances. $\alpha$ and $\beta$ are the punishment and reward factors, respectively. $l$ denotes the length of emission interval, where there are $2K$ intervals, $k=1, 2, 3, \cdots, K$. {When the difference between the carbon emission quota and the actual carbon emissions does not exceed $l$, the price of carbon trading is $p^{\rm{c}}$. Thereafter, whenever the difference between carbon emission quota and actual carbon emissions exceeds an interval length $l$, the carbon trading price for the excess part will increase by $\alpha p^{\rm{c}}$. For example, when the actual carbon emission of IC $i$ exceeds its carbon emission quota by $1.4l$, the price of trading $l$ carbon emission right is $p^{\rm{c}}$, and the price of carbon right for the excess $0.4l$ is $(1+\alpha)p^{\rm{c}}$. }% $Q_{ia}$ is the actual carbon emission of IC $i$.

\subsection{Multi-Energy Trading}
To satisfy multi-energy demand, ICs in the industrial park can trade gas and electricity with each other. The deal energy price and quantity are calculated by the multi-energy trading and scheduling method, which is elaborated in Section III. %Firstly, the sellers and buyers among ICs decide the energy price and quantity based on cost minimization, and then send the energy price and quantity to the auctioneer. Secondly, according the deal electricity price and quantity, the price and quantity of gas trading are determined. Finally, the quantity of carbon trading is determined.

%The heat operation mode under variable supply temperature and constant flow is adopted, which decouples the thermal conditions and the control of hydraulic regimes and is suitable for practical industrial production \cite{Li2016Transmission}. Under this mode, the pumps have fixed energy consumption. There 

According to the deal energy price and quantity, the cost $C_{i}^{\rm{X}}(t)$ and revenue $R_{i}^{\rm{S}}(t)$  of IC $i$ are 
%\begin{subequations}
\begin{equation}
C_{i}^{\rm{X}}(t)=\sum_{j}p_{ij}^{\rm{e}}(t)X_{ij}^{\rm{eb}}(t)+\sum_{j}p_{ij}^{\rm{g}}(t)X_{ij}^{\rm{gb}}(t), \forall i,t
\label{cfpq}
\end{equation}
\begin{equation}
R_{i}^{\rm{S}}(t)=\sum_{j}s_{ij}^{\rm{e}}(t)X_{ij}^{\rm{es}}(t)+\sum_{j}s_{ij}^{\rm{g}}(t)X_{ij}^{\rm{gs}}(t), \forall i,t
\label{rfpq}
\end{equation}
%\begin{tcolorbox}
%\end{tcolorbox}
%\begin{equation}
% X^2_{ijgs}(t)=C^2_{ij}(\pi^2_{i}(t)-\pi^2_{j}(t))
% \label{weco}
%\end{equation}
%\begin{equation}
%\pi_{min}\leq \pi_i(t)\leq \pi_{max}
%\end{equation}
%\begin{equation}
%\rho_{min}\leq \frac{\pi_i(t)}{\pi_j(t)}\leq \rho_{max}
%\end{equation}
%\label{depq}
%\end{subequations}
where $p_{ij}^{\rm{e}}(t)$ and $p_{ij}^{\rm{g}}(t)$ are the electricity and gas prices that buyer $i$ purchases from seller $j$. $X_{ij}^{\rm{eb}}(t)$ and $X_{ij}^{\rm{gb}}(t)$ denote the quantities of energy purchased by buyer $i$ from seller $j$. $s_{ij}^{\rm{e}}(t)$ and $s_{ij}^{\rm{g}}(t)$ denote the energy prices sold by seller $i$ to buyer $j$. $X_{ij}^{\rm{es}}(t)$ and $X_{ij}^{\rm{gs}}(t)$ are the quantities of energy sold by seller $i$ to buyer $j$. This paper mainly considers the influence of node pressure on gas flow in natural gas trading. {$\forall ij \in \Omega_c$ with $\Omega_c$ being the set of ICs that trade energy with each other.} The gas purchased by IC $i$ from the node $n$ of IC $j$ is delivered to the node $m$ of IC $i$ through the pipeline $nm$ 
\begin{equation}
X_{ij}^{\rm{gb}}(t)=f_{nm}(t), \forall ij,nm,t
\label{11d}
\end{equation}

%$X_{ieb}(t)$ and $X_{igb}(t)$ denote the purchases of electricity and gas. $X_{ies}(t)$ and $X_{igs}(t)$ denote sales of electricity and gas. %$X_{ijgb}$ and $X_{ijgs}$ denote the gas purchased and sold from industrial cluster $i$ to $j$. $C_{ij}$ is the Weymouth constant of line $ij$. $\pi_{i}$ and $\pi_{j}$ denote the gas pressure of node $i$ and $j$.

When the energy in the trading market among ICs is insufficient, IC $i$ will purchase electricity $E_i(t)$ and gas $G_i(t)$ from the CCPP and natural gas plant. When the renewable energy generation is surplus, IC $i$ will sell electricity $E_{i}^{\rm{o}}(t)$ to the CCPP. The constraints on energy trading with plants are 
\begin{equation}
\begin{aligned}
&0\leq E_i(t) \leq E_{i}^{\max},  0\leq G_i(t) \leq G_{i}^{\max}, \\& 0\leq E_{i}^{\rm{o}}(t) \leq E_{i}^{\rm{o,max}}, \forall i,t
\label{eo2}
\end{aligned}
\end{equation}
\subsection{Energy Demand}
%The total available energy for industrial users depends on the energy flows of MEGPs and utility companies. 
{The total electricity $E_{i}^{\rm{d}}(t)$, gas $G_{i}^{\rm{d}}(t)$ and heat $H_{i}^{\rm{d}}(t)$ demands of the users in IC $i$ are satisfied by multi-energy devices and utility companies. } %total available energy can be denoted as:
\begin{equation}
\begin{aligned}
E_{i}^{\rm{d}}(t)&=E_{i}(t)-E_{i}^{\rm{o}}(t)+X_{i}^{\rm{eb}}(t)-X_{i}^{\rm{es}}(t)+D_{i}^{\rm{e}}(t)\\&-C_{i}^{\rm{e}}(t)+E_{i}^{\rm{CHP}}(t)-E_{i}^{\rm{p2g}}(t)+E_{i}^{\rm{r}}(t), \forall i,t
\label{11a}
\end{aligned}
\end{equation}
\begin{equation}
\begin{aligned}
 G_{i}^{\rm{d}}(t)&=G_{i}(t)-G_{i}^{\rm{CHP}}(t)-G_{i}^{\rm{b}}(t)+X_{i}^{\rm{gb}}(t)-X_{i}^{\rm{gs}}(t)\\&+G_{i}^{\rm{p2g}}(t), \forall i,t
 \label{11b}
\end{aligned}
 \end{equation}
\begin{equation}
\begin{aligned}
H_{i}^{\rm{d}}(t)=H_{i}^{\rm{CHP}}(t)+H_{i}^{\rm{b}}(t)+D_{i}^{\rm{h}}(t)-C_{i}^{\rm{h}}(t), \forall i,t
\label{11c}
\end{aligned}
\end{equation}
\begin{equation}
%\text{s.t. } 
%\mathcolorbox{yellow}
{X_{i}^{\rm{eb}}(t)=\sum_jX_{ij}^{\rm{eb}}(t), X_{i}^{\rm{es}}(t)=\sum_jX_{ij}^{\rm{es}}(t)}, \forall i,t
\end{equation}
\begin{equation}
%\text{s.t. } 
%\mathcolorbox{yellow}
{X_{i}^{\rm{gb}}(t)=\sum_jX_{ij}^{\rm{gb}}(t), X_{i}^{\rm{gs}}(t)=\sum_jX_{ij}^{\rm{gs}}(t)}, \forall i,t
\end{equation}
%where $E_{it}(t)$, $G_{it}(t)$ and $H_{it}(t)$ denote the total electricity, gas and heat demand of IC $i$, respectively. 

Gas is not used directly, but converted into heat and electricity to satisfy users' demands, so $G_{i}^{\rm{d}}(t)$ is 0. %In addition, the energy supply and demand should be balanced in real time. 
%$E_{ir}(t)$ is the renewable energy generation. +O_{ic}(t)p_{co}(t) and and $p_{co}(t)$ coal price
\subsection{Cost Function}
The total cost of IC $i$ consists of the trading cost and revenue 
\begin{equation}
\begin{aligned}
C_{i}(t)&=C_{i}^{\rm{c}}(t)+E_{i}(t)p^{\rm{e}}(t)-E_{i}^{\rm{o}}(t)p^{\rm{o}}(t)\\&+[G_{i}^{\rm{CHP}}(t)+G_{i}^{\rm{b}}(t)]p^{\rm{g}}(t)+C_{i}^{\rm{X}}(t)-R_{i}^{\rm{S}}(t), \forall i,t
\label{tcic}
\end{aligned}
\end{equation}
where $p^{\rm{e}}(t)$ denotes the electricity price of the CCPP and $p^{\rm{g}}(t)$ denotes the natural gas price. $p^{\rm{o}}(t)$ denotes the electricity price sold back to the CCPP. %$p_{g}(t)$ denotes the gas price of the gas plant.

The aim of IC $i$ is to minimize the total cost with the time-varying characteristics of energy supply and demand  %cost through carbon and energy trading and scheduling
\begin{align}
\min_{\boldsymbol{M}_i(t)}\lim_{T\rightarrow\infty}\frac{1}{T}\sum_{t=0}^{T-1}\mathbb{E}\{C_i(t)\}, \forall i
\label{eq301}
\end{align}
\begin{equation}
\nonumber
\text{s.t. } (\ref{A1}) - (\ref{11c})
\end{equation}
where $\boldsymbol{M}_i(t)$=\{$D_{i}^{\rm{e}}(t)$, $C_{i}^{\rm{e}}(t)$, $D_{i}^{\rm{h}}(t)$, $C_{i}^{\rm{h}}(t)$, $E_{i}^{\rm{o}}(t)$, $E_i(t)$, $E_{i}^{\rm{p2g}}(t)$, $G_{i}^{\rm{CHP}}(t)$, $G_{i}^{\rm{b}}(t)$, $X_{i}^{\rm{eb}}(t)$, $X_{i}^{\rm{gb}}(t)$, $X_{i}^{\rm{es}}(t)$, $X_{i}^{\rm{gs}}(t)$, {$\forall i,t$}\} denotes the set of optimization variables.

\section{Solution Method}
{Different from traditional methods like dynamic programming, which requires priori information of the random processes in the system and has high computational complexity, Lyapunov optimization can give simple online solutions according to the current information about the system state. The performance of the Lyapunov optimization-based algorithm can be arbitrarily close to the optimal performance }\cite{Lakshminarayana2014Cooperation}. {Fundamental assumptions about future information unavailability make offline methods impractical for systems with a high degree of uncertainty, and dynamic programming is unsuitable for multiple networked energy storage systems }\cite{Sarthak2018Optimal}.

Firstly, Lyapunov optimization is introduced to convert the original problem into a weighted cost minimization problem for each slot, which can make control decisions to weigh the overall energy cost and energy storage stability. 
Then, a bound tightening algorithm is adopted to handle the  relaxation problem of the quadratic constraint on gas transmission among ICs. Finally, the energy trading among ICs is determined by matching game.
\subsection{Lyapunov optimization}
In order to handle the coupling constraints of electricity and heat charging/discharging, the time-average constraints of the relaxed problem are used to replace the coupling constraints (\ref{A1}) and (\ref{A3}) of the original problem
\begin{equation}
\begin{split}
\lim_{T\rightarrow\infty}\frac{1}{T}\sum_{t=0}^{T-1}\mathbb{E}\{C_{i}^{\rm{e}}(t)\}=\lim_{T\rightarrow\infty}\frac{1}{T}\sum_{t=0}^{T-1}\mathbb{E}\{D_{i}^{\rm{e}}(t)\}, \forall i\\
\lim_{T\rightarrow\infty}\frac{1}{T}\sum_{t=0}^{T-1}\mathbb{E}\{C_{i}^{\rm{h}}(t)\} =\lim_{T\rightarrow\infty}\frac{1}{T}\sum_{t=0}^{T-1}\mathbb{E}\{D_{i}^{\rm{h}}(t)\}, \forall i\\
\end{split}
\label{sto1}
\end{equation}

{$C_{i}^{\rm{opt}}$ denotes the optimal cost of the original problem and $C_{ir}^{\rm{opt}}$ denotes the optimal cost of the relaxed problem. Since the relaxed problem has fewer constraints than the original problem, $C_{ir}^{\rm{opt}} \leq C_{i}^{\rm{opt}}$ holds.}

{The stationary and randomized policy $\Pi$ is introduced to obtain the optimal solution of the relaxed problem}
\begin{equation}
\begin{aligned}
\mathbb{E}\{C_{i}^{\Pi}(t)\}=C_{ir}^{\rm{opt}}, \forall i,t
\end{aligned}
\end{equation}
subject to: 
\begin{equation}
\begin{split}
C^{\rm{e},\Pi}_{i}(t)& = D^{\rm{e},\Pi}_{i}(t), \forall i,t\\
%C^{\rm{y},\Pi}_{i}(t) &= D^{\Pi}_{iy}(t)\\
C^{\rm{h},\Pi}_{i}(t) &= D^{\rm{h},\Pi}_{i}(t), \forall i,t\\
%D^{\Pi}_{iyl}(t)+d^{\Pi}_{il}(t) &= Y^{\Pi}_{ifl}(t)+h^{\Pi}_{il}(t)\\
\end{split}
\label{cd2}
\end{equation} and  (\ref{Cem}) - (\ref{11c}).

The policy $\Pi$ can be ensured by the Caratheodory theory, similar to \cite{Leonidas2006Resource}. 
 Clearly, solutions to the relaxed problem are feasible to the original problem only if they satisfy the constraint (\ref{Wm}). 
To achieve the goal, virtual energy storage queues $F_{i}(t)$ and $Z_{i}(t)$ are constructed as $F_{i}(t)=B_{i}(t)-\theta_{i}, Z_{i}(t)=W_{i}(t)- \varepsilon_{i}$, $\forall i,t$. 
%\begin{equation}
%\begin{aligned}
%F_{i}(t)=B_{i}(t)-\theta_{i}, Z_{i}(t)=W_{i}(t)- \varepsilon_{i}
%\end{aligned}
%\label{vir}
%\end{equation}
$\theta_{i}$ and $\varepsilon_{i}$ are utilized to ensure $B_{i}(t)$ and $W_{i}(t)$ satisfy the upper limits \cite{Zhu2020Energy}. 

The Lyapunov function and Lyapunov drift are
\begin{equation}
Y_{i}(t)=\frac{1}{2}F_{i}(t)^{2}+\frac{1}{2}Z_{i}(t)^{2}, \forall i,t
\label{lf}
\end{equation}
\begin{equation}
\Delta_{i}(t)=\mathbb{E}\{Y_{i}(t+1)-Y_{i}(t)|B_{i}(t),W_{i}(t)\}, \forall i,t
\end{equation}
{where the expectation is with respect to the stochastic process of the system, given $B_{i}(t)$ and $W_{i}(t)$. }
According to the virtual queue $F_{i}(t)$ and $Z_{i}(t)$, the Lyapunov drift has the upper bound
\begin{equation}
\begin{aligned}
\Delta_{i}(t)%&=\mathbb{E}\{Q_{i}(t+1)-Q_{i}(t)|B_{i}(t),W_{i}(t)\} \\
&\leq A_{i}+\mathbb{E}\{F_{i}(t)(C_{i}^{\rm{e}}(t)-D_{i}^{\rm{e}}(t))\\&+Z_{i}(t)(C_{i}^{\rm{h}}(t)-D_{i}^{\rm{h}}(t))\}, \forall i,t\\
A_{i}&=\frac{1}{2}[\max((C_{i}^{\rm{e,max}})^2,(D_{i}^{\rm{e,max}})^2)\\&+\max((C_{i}^{\rm{h,max}})^2,(D_{i}^{\rm{h,max}})^2)]
\end{aligned}
\label{ldub}
\end{equation}
%where $A_{i}$ is a constant and
%where $A_{i}=\frac{1}{2}[\max((C_{i}^{\rm{e,max}})^2,(D_{i}^{\rm{e,max}})^2)+\max((C_{i}^{\rm{h,max}})^2,(D_{i}^{\rm{h,max}})^2)]$. 

The proof of this step can refer to \cite{Zhu2020Energy}.

To minimize overall cost of the IC while ensuring energy storage stability, $V_i$ is used to control the tradeoff between energy cost and queue stability. The corresponding drift-plus-penalty term is
\begin{equation}
\begin{aligned}
&\Delta_{i}(t)+V_{i}\mathbb{E}\{C_{i}(t)\} \\&\leq A_{i}+\mathbb{E}\{F_{i}(t)(C_{i}^{\rm{e}}(t)-D_{i}^{\rm{e}}(t))
+Z_{i}(t)(C_{i}^{\rm{h}}(t)-D_{i}^{\rm{h}}(t))
\\&+V_{i}(C_{i}^{\rm{c}}(t)+E_{i}(t)p^{\rm{e}}(t)-E_{i}^{\rm{o}}(t)p^{\rm{o}}(t)+C_{i}^{\rm{X}}(t)-R_{i}^{\rm{S}}(t)\\&+(G_{i}^{\rm{CHP}}(t)+G_{i}^{\rm{b}}(t))p^{\rm{g}}(t))\}, \forall i,t
\end{aligned}
\label{p1}
\end{equation}

The original optimization problem (\ref{eq301}) is converted to minimize $\Gamma_{i}(t)$
\begin{equation}
\begin{aligned}
&\min_{\boldsymbol{M}_{i}(t)}  \Gamma_{i}(t)\\&=\min_{\boldsymbol{M}_{i}(t)}  F_{i}(t)(C_{i}^{\rm{e}}(t)-D_{i}^{\rm{e}}(t))
+Z_{i}(t)(C_{i}^{\rm{h}}(t)-D_{i}^{\rm{h}}(t))
\\&+V_{i}(C_{i}^{\rm{c}}(t)+E_{i}(t)p^{\rm{e}}(t)-E_{i}^{\rm{o}}(t)p^{\rm{o}}(t)+C_{i}^{\rm{X}}(t)-R_{i}^{\rm{S}}(t)\\&+(G_{i}^{\rm{CHP}}(t)+G_{i}^{\rm{b}}(t))p^{\rm{g}}(t)), \forall i,t
\end{aligned}
\label{p1}
\end{equation}
\begin{equation}
\nonumber
\text{s.t. } (\ref{Cem}) - (\ref{11c})%, (\ref{sto1}) 
\end{equation}

\subsection{Constraints Relaxation of Gas flows}
Owing to the quadratic constraint (\ref{weco}) on gas transmission among ICs, the system model is nonlinear. To deal with the nonlinear problem, the quadratic constraint (\ref{weco}) is replaced by
\begin{equation}
\frac{f^2_{nm}(t)}{C^2_{nm}}\geq \pi^2_{n}(t)-\pi^2_{m}(t), \forall nm,t
 \label{qcr1}
 \end{equation}
 \begin{equation}
 \frac{f^2_{nm}(t)}{C^2_{nm}}\leq \pi^2_{n}(t)-\pi^2_{m}(t), \forall nm,t
 \label{qcr2}
\end{equation}

After relaxing (\ref{qcr1}) and replacing (\ref{weco}) with (\ref{qcr2}), the trading and scheduling optimization problem can be solved with second-order cone programming. However, the relaxed constraints may lead to serious violations of (\ref{weco}). {Thus, a relaxation method based on quadratic convex relaxation} \cite{Chen2019Unit} {is introduced to guarantee the approximation of the constraint }(\ref{weco}). To avoid the extra computation and less tight relaxation caused by individually approximating $\pi^{2}_{n}(t)$ and $\pi^{2}_{m}(t)$, the pressure difference $\pi_{nm}^{\rm{d}}(t)$ and the pressure sum $\pi_{nm}^{\rm{s}}(t)$ between nodes $n$ and $m$ are introduced 
\begin{equation}
\pi_{nm}^{\rm{d}}(t)= \pi_{n}(t)-\pi_{m}(t), \forall nm,t
 \label{dnm}
 \end{equation}
 \begin{equation}
 \pi_{nm}^{\rm{s}}(t)= \pi_{n}(t)+\pi_{m}(t), \forall nm,t
 \label{snm}
\end{equation}

\begin{figure}
  \centering
  \includegraphics[width=.8\hsize]{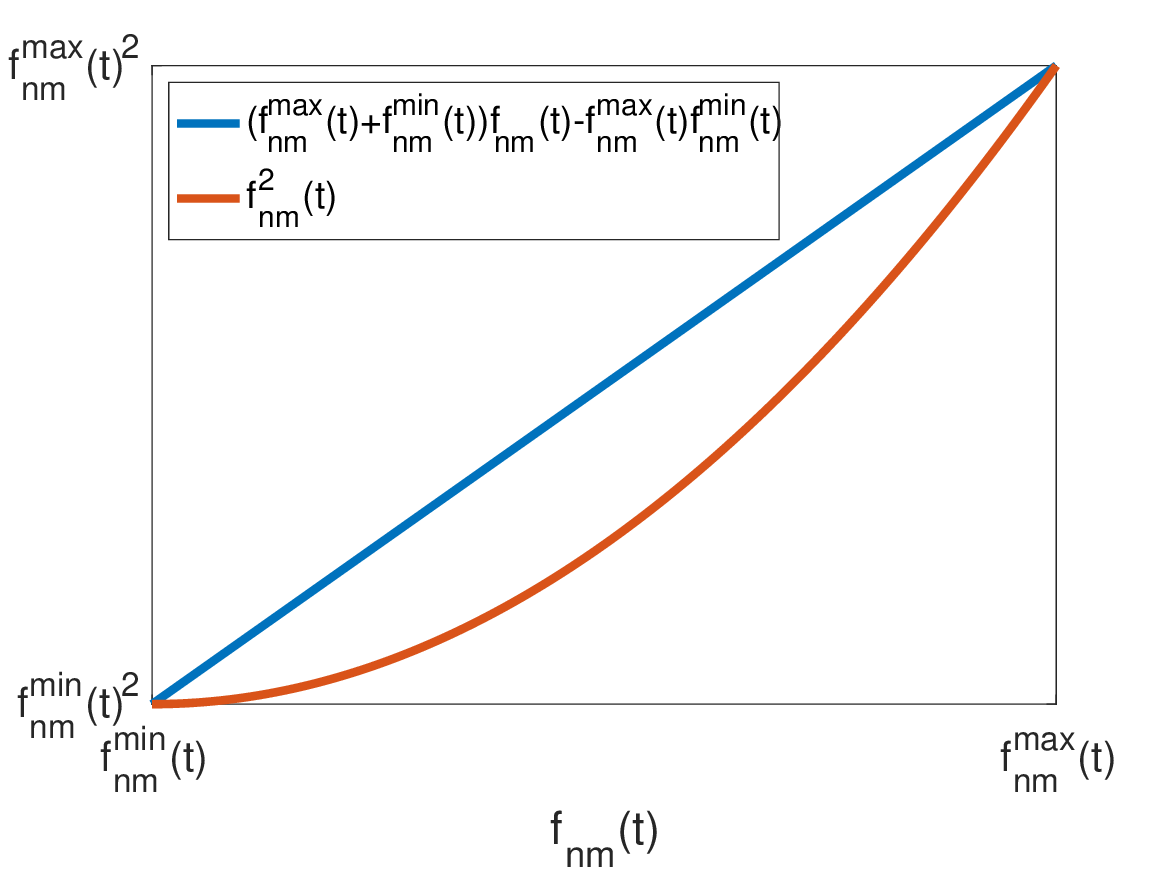}
  \caption{Approximation of $f^{2}_{nm}(t)$ according to (\ref{tr2}) and (\ref{tr3}).}
  \label{f4}
\end{figure}

The relaxation of (\ref{qcr1}) is denoted as
\begin{equation}
\frac{F_{nm}(t)}{C^2_{nm}}\geq \Pi_{nm}(t), \forall nm,t
 \label{tr1}
 \end{equation}
 \begin{equation}
 \begin{aligned}
F_{nm}(t) \geq f^2_{nm}(t), \forall nm,t
 \label{tr2}
\end{aligned}
\end{equation}
\begin{equation}
\begin{aligned}
F_{nm}(t) \leq (f_{nm}^{\max}(t)+f_{nm}^{\min}(t))f_{nm}(t)-f_{nm}^{\max}(t)&f_{nm}^{\min}(t), \\&\forall nm,t
\end{aligned}
 \label{tr3}
\end{equation}
 \begin{equation}
\begin{aligned}
\Pi_{nm}(t) \geq \pi_{nm}^{\rm{d},\min}(t)(\pi_{nm}^{\rm{s}}(t)-\pi_{nm}^{\rm{s},\min}(t))+\pi_{nm}^{\rm{s},\min}(t)&\pi_{nm}^{\rm{d}}(t), \\&\forall nm,t
\end{aligned}
 \label{tr4}
\end{equation}
 \begin{equation}
\begin{aligned}
\Pi_{nm}(t) \geq \pi_{nm}^{\rm{d},\max}(t)(\pi_{nm}^{\rm{s}}(t)-\pi_{nm}^{\rm{s},\max}(t) 
)+\pi_{nm}^{\rm{s},\max}(t)&\pi_{nm}^{\rm{d}}(t), \\&\forall nm,t
\end{aligned}
\label{tr5}
\end{equation}
 \begin{equation}
\begin{aligned}
\Pi_{nm}(t) \leq \pi_{nm}^{\rm{d},\min}(t)(\pi_{nm}^{\rm{s}}(t)-\pi_{nm}^{\rm{s},\max}(t))+\pi_{nm}^{\rm{s},\max}(t)&\pi_{nm}^{\rm{d}}(t), \\&\forall nm,t
\end{aligned}
 \label{tr6}
\end{equation}
 \begin{equation}
\begin{aligned}
\Pi_{nm}(t) \leq \pi_{nm}^{\rm{d},\max}(t)(\pi_{nm}^{\rm{s}}(t)-\pi_{nm}^{\rm{s},\min}(t))+\pi_{nm}^{\rm{s},\min}(t)&\pi_{nm}^{\rm{d}}(t), \\&\forall nm,t
\end{aligned}
 \label{tr7}
\end{equation}
where $F_{nm}(t)$ and $\Pi_{nm}(t)$ denote approximations of $f^{2}_{nm}(t)$ and $\pi^2_{n}(t)-\pi^2_{m}(t)$ to achieve the replacement of (\ref{qcr1}) with (\ref{tr1}). Fig.~\ref{f4} shows that convex boundary constraints (\ref{tr2})-(\ref{tr3}) of $f^2_{nm}(t)$ impose the bounds on $F_{nm}(t)$. Likewise, convex boundary constraints (\ref{tr4})-(\ref{tr7}) of $\pi^{2}_{n}(t)-\pi^{2}_{m}(t)$ (i.e., $\pi_{nm}^{\rm{d}}(t)\pi_{nm}^{\rm{s}}(t)$) impose similar bounds on $\Pi_{nm}(t)$. %According to the definition of $\pi_{dnm}(t)$ and $\pi_{snm}(t)$, $\pi_{dnm}(t)\pi_{snm}(t)=\pi^{2}_{n}(t)-\pi^{2}_{m}(t)$. Thus, $\Pi_{nm}(t)$ is related to $\pi_{dnm}(t)$ and $\pi_{snm}(t)$. Convex boundary constraints (\ref{tr4})-(\ref{tr7}) of $\pi_{dnm}(t)\pi_{snm}(t)$ impose the bounds on $\Pi_{nm}(t)$. 
The optimization problem (\ref{p1}) is converted to the relaxed problem 
 \begin{align}
\min_{\boldsymbol{M}_{i}(t)}  \Gamma_{i}(t), \forall i,t
\label{eq302}
\end{align}
\begin{equation}
\nonumber
\text{s.t. } (\ref{Cem}) - (\ref{ncig}), (\ref{ipa}) - (\ref{11c}), %(\ref{sto1}), 
(\ref{qcr2}) - (\ref{tr7})
\end{equation}

Fig.~\ref{f4} shows that the approximate effect of the constraints (\ref{tr2})-(\ref{tr3}) depends on the values $f_{nm}^{\min}(t)$, $f_{nm}^{\max}(t)$. For instance, the maximum deviation between $f^{2}_{nm}(t)$ and $(f_{nm}^{\max}(t)+f_{nm}^{\min}(t))f_{nm}(t)-f_{nm}^{\max}(t)f_{nm}^{\min}(t)$ is $[(f_{nm}^{\max}(t)-f_{nm}^{\min}(t))/2]^2$. Similarly, the approximate effect of the constraints (\ref{tr4})-(\ref{tr7}) depends on the values $\pi_{nm}^{\rm{d},\min}(t)$, $\pi_{nm}^{\rm{d},\max}(t)$, $\pi_{nm}^{\rm{s},\min}(t)$ and $\pi_{nm}^{\rm{s},\max}(t)$.

%\begin{CJK}{UTF8}{gbsn}
%\begin{algorithm}[h]
%\floatname{algorithm}{算法}
%\renewcommand{\algorithmicrequire}{\textbf{Communication}}
%\renewcommand{\algorithmicensure}{\textbf{Contribution}}
%\footnotesize
%\caption{: 边界紧缩算法}
%\label{alg1}
%\begin{algorithmic}[1]
%  \State \textbf{Initialize:} $f_{nm}^{min}(t)$, $f_{nm}^{max}(t)$, $\pi_{dnm}^{min}(t)$, $\pi_{dnm}^{max}(t)$, $\pi_{snm}^{min}(t)$ 和 $\pi_{snm}^{max}(t)$
%    \For{$k = 1, 2, ..., K$}
%\State 解决松弛优化问题得到 $f_{nm}^{k}(t)$, $\pi_{dnm}^{k}(t)$ 和 $\pi_{snm}^{k}(t)$
%\State $f^{min}_{nm}(t) \leftarrow (1-\sigma_k)f_{nm}^{k}(t)$, $f^{max}_{nm}(t) \leftarrow (1+\sigma_k)f_{nm}^{k}(t)$
%\State $\pi^{min}_{dnm}(t) \leftarrow (1-\sigma_k)\pi_{dnm}^{k}(t)$, $\pi^{max}_{dnm}(t) \leftarrow (1+\sigma_k)\pi_{dnm}^{k}(t)$
%\State $\pi^{min}_{snm}(t) \leftarrow (1-\sigma_k)\pi_{snm}^{k}(t)$, $\pi^{max}_{snm}(t) \leftarrow (1+\sigma_k)\pi_{snm}^{k}(t)$
%\State \textbf{Until} $|\pi^{2}_{n}(t)-\pi^{2}_{m}(t)-f^{2}_{nm}(t)/C^{2}_{nm}|\leq \delta \pi^{2}_{n}(t)$
%\EndFor 
%\label{code:recentEnd}
%\end{algorithmic}
%\end{algorithm}
%\end{CJK}

\begin{algorithm}[h]
\floatname{algorithm}{Algorithm}
\renewcommand{\algorithmicrequire}{\textbf{Communication}}
\renewcommand{\algorithmicensure}{\textbf{Contribution}}
\footnotesize
\caption{: Bound Tightening Algorithm}
\label{alg1}
\begin{algorithmic}[1]
  \State \textbf{Initialize:} $f_{nm}^{\min}(t)$, $f_{nm}^{\max}(t)$, $\pi_{nm}^{\rm{d},\min}(t)$, $\pi_{nm}^{\rm{d},\max}(t)$, $\pi_{nm}^{\rm{s},\min}(t)$, $\pi_{nm}^{\rm{s},\max}(t)$.   
    \For{$o = 1, 2, \cdots, O$}
\State Solve the relaxed problem to obtain $f_{nm,o}(t)$, $\pi_{nm,o}^{\rm{d}}(t)$ and $\pi_{nm,o}^{\rm{s}}(t)$.
\State $f^{\min}_{nm}(t) \leftarrow (1-\sigma_o)f_{nm,o}(t)$, $f^{\max}_{nm}(t) \leftarrow (1+\sigma_o)f_{nm,o}(t)$.
\State $\pi^{\rm{d},\min}_{nm}(t) \leftarrow (1-\sigma_o)\pi_{nm,o}^{\rm{d}}(t)$, $\pi^{\rm{d},\max}_{nm}(t) \leftarrow (1+\sigma_o)\pi_{nm,o}^{\rm{d}}(t)$.
\State $\pi^{\rm{s},\min}_{nm}(t) \leftarrow (1-\sigma_o)\pi_{nm,o}^{\rm{s}}(t)$, $\pi^{\rm{s},\max}_{nm}(t) \leftarrow (1+\sigma_o)\pi_{nm,o}^{\rm{s}}(t)$.
\State \textbf{Until} $|\pi^{2}_{n}(t)-\pi^{2}_{m}(t)-f^{2}_{nm}(t)/C^{2}_{nm}|\leq \delta \pi^{2}_{n}(t)$.
\EndFor 
\label{code:recentEnd}
\end{algorithmic}
\end{algorithm}

Therefore, a bound tightening algorithm is presented to enhance the tightness of the proposed relaxation. Algorithm 1 shows the process of the bound tightening algorithm, where the variable $\sigma_o$ is gradually reduced to ensure that the upper and lower bounds converge within the tight range where the optimal solution lies, and $\delta$ is the convergence threshold.

\subsection{Matching Game}
%Owing to the unbalance existence of energy supply and demand in individual ICs, energy needs to be shared among ICs. First, the selling price $\widetilde{\alpha}_{i}(t)$ and purchase price $\widetilde{\beta}_{i}(t)$ of IC $i$  are determined according to \cite{Zhu2020Energy}: $\widetilde{\alpha}_{i}(t)=\max[\frac{-A_{i}(t)}{V_{i}},\frac{-F_{i}(t)}{(\frac{h}{\eta_{e}}+c_1)V_{i}}, p_{eo}(t)], \widetilde{\beta}_{i}(t)=\min[\frac{\max(-A_{i}(t),0)}{V_{i}},\frac{p_{g}}{\eta_{pg}},p_{e}(t)]$. Then, the amount of energy sold by sellers and purchased by buyers are determined by solving the optimization problem (\ref{eq302}).

Due to energy supply and demand imbalance in individual ICs, energy needs to be shared among ICs. First, the submitted selling price or purchase price of IC $i$ is determined in energy trading according to \cite{Zhu2020Energy}. Then, the amount of energy sold by sellers and purchased by buyers are obtained by solving the problem (\ref{eq302}). The two sides of the transaction can be determined by the matching game.

The matching theory based on the preference and information of ICs provides low-complexity distributed solutions to the energy matching problem between sellers and buyers. Thus, a many-to-many matching game model is introduced, where a seller can sell energy to several buyers and a buyer can purchase energy from several sellers. $I$ and $J$ respectively denote the set of buyers and sellers in the trading market. A matching game $\mu$ is defined as a mapping from set $I \cup J$ into all subsets of set $I \cup J$, where 1) $\forall i \in I$, $\mu(i) \in J$; 2) $\forall j \in J$, $\mu(j) \in I$; 3) if and only if $\mu(i) =j$, $\mu(j) =i$.

%For any buyer $i\in I$, it will choose seller $j$ that can maximize the reduction of the objective function in (\ref{eq302}). Therefore, the utility of buyer $i$ is denoted as $U_{i}^{I}(j)=\gamma_i{(J_i \cup j)}-\gamma_i(J_i)$, where $\gamma_i(J_i)=\Gamma_i(t)$ and $J_i$ is the set of sellers matched by buyer $i$. Similarly, the utility of seller $j$ is denoted as $U_{j}^{J}(i)=\gamma_{(I_j \cup i)}(t)-\gamma(I_j)$.

For any buyer $i\in I$, it will choose seller $j$ that can maximize its energy trading revenue. The revenue of buyer $i$ is denoted as $U_{i}(j)=-C_{i}^{\rm{X}}(t)=-\sum_{j}p_{ij}^{\rm{e}}(t)X_{ij}^{\rm{eb}}(t)-\sum_{j}p_{ij}^{\rm{g}}(t)X_{ij}^{\rm{gb}}(t),\forall i,t$. Similarly, the revenue of seller $j$ is denoted as $U_{j}(i)=\sum_{i}s_{ij}^{\rm{e}}(t)X_{ij}^{\rm{es}}(t)+\sum_{i}s_{ij}^{\rm{g}}(t)X_{ij}^{\rm{gs}}(t), \forall j,t$. The deal energy price is the average of matching bid and ask quotes. 

A preference relation $\succ$ is introduced to rank the preference list of buyers and sellers. $\succ_{i}$ of buyer $i$ is used to rank sellers over the set of $J$, where $j \succ_{i} j' \Leftrightarrow U_{i}(j)>U_{i}(j')$. Similarly, $i \succ_{j} i' \Leftrightarrow U_{j}(i)>U_{j}(i')$.

With the preference relations and the utility functions of buyers and sellers, the energy trading problem between buyers and sellers is solved by the matching game based energy trading algorithm, as shown in Algorithm 2.  The algorithm 2 can converge to a stable matching  \cite{Hatfield2017Many}, i.e., there does not exist a buyer $i' \in I$ such that $i' \succ_{j}\mu(j), \forall j$. After determining the energy trading among ICs, the optimal solution set $\boldsymbol{M}^*_{i}(t)$ of each IC can be obtained by solving the problem (\ref{eq302}). The whole procedure of the proposed method is shown in Algorithm 3.

\begin{algorithm}[h]
\floatname{algorithm}{Algorithm}
\renewcommand{\algorithmicrequire}{\textbf{Communication}}
\renewcommand{\algorithmicensure}{\textbf{Contribution}}
\footnotesize
\caption{: Matching Game Based Trading Algorithm}
\label{alg1}
\begin{algorithmic}[1]
    \While{unstable matching}
%\State \textbf{Stage I:} for each buyer 
%\State  Calculate each buyer's utility values about all sellers that are not matched in current iteration according to $U_{i}(j)$.
\State  Update each buyer's preference list about sellers according to $U_{i}(j)$, and send its intention to its most preferred seller.
%\State \textbf{Stage II:} for each seller
%\State Each Calculate the seller's utility values about buyers that send intentions to it and the buyers that matched the seller according to $U_{j}(i)$.
\State  Update each seller's preference list about the intentional and matched buyers according to $U_{j}(i)$.
\State \textbf{repeat:} 
\State  Randomly choose a seller $j$ to sell its energy to its most preferred buyers. If the energy of $j$ was matched by other buyer $i'$, remove the energy of $j$ from the matching energy of $i'$.
\State \textbf{until} The seller sells all energy or all preferred buyers are matched.
\State When a buyer buys the energy it needs, remove it from $I$. \State When a seller sells all energy, remove it from $J$. 
\EndWhile 
\label{code:recentEnd}
\end{algorithmic}
\end{algorithm}

\begin{algorithm}[h]
\floatname{algorithm}{Algorithm}
\renewcommand{\algorithmicrequire}{\textbf{Communication}}
\renewcommand{\algorithmicensure}{\textbf{Contribution}}
\footnotesize
\caption{: {The Procedure of the Proposed Method}}
\label{alg1}
\begin{algorithmic}[1]
%\State Set $t=0$.
\State \textbf{Initialize:} $B_{i}(t)$, $W_{i}(t)$.
\For{IC $i$}
\State Calculate the submitted selling and purchase prices of IC $i$, and calculate $X_{i}^{\rm{eb}}(t)$, $X_{i}^{\rm{gb}}(t)$, $X_{i}^{\rm{es}}(t)$ and $X_{i}^{\rm{gs}}(t)$ by (\ref{eq302}) based on Lyapunov optimization and the bound tightening algorithm.
\State Calculate $p_{ij}^{\rm{e}}(t)$, $s_{ij}^{\rm{e}}(t)$, $p_{ij}^{\rm{g}}(t)$, $s_{ij}^{\rm{g}}(t)$, $X_{ij}^{\rm{eb}}(t)$, $X_{ij}^{\rm{gb}}(t)$, $X_{ij}^{\rm{es}}(t)$ and $X_{ij}^{\rm{gs}}(t)$ by the matching game.
\State Calculate $\boldsymbol{M}_i(t)$ using (\ref{eq302}).
\EndFor
\State Calculate $B_{i}(t+1)$ and $W_{i}(t+1)$ by (\ref{A1}) - (\ref{A3}).
\label{code:recentEnd}
\end{algorithmic}
\end{algorithm}

\subsection{Performance Analysis}
In the above methods, we consider the battery and water tank of IC $i$ as follows

\textbf{Lemma 1.} {When $\theta_i$, $\epsilon_i$ and $V_i$ are set as }
\begin{equation}
\begin{aligned}
\theta_i&=V_ip^{\rm{e,max}}+D_{i}^{\rm{e,max}}\\
\epsilon_i&=\frac{V_ip^{\rm{g,max}}}{\eta_{i}^{\rm{bg}}}+D_{i}^{\rm{h,max}}\\
 V_{i}^{\rm{max}}&=\min\{\frac{B_{i}^{\rm{max}}-C_{i}^{\rm{e,max}}-D_{i}^{\rm{e,max}}}{p^{\rm{e,max}}},\\&\frac{\eta_{i}^{\rm{bg}}(W_{i}^{\rm{max}}-C_{i}^{\rm{h,max}}-D_{i}^{\rm{h,max}})}{p^{\rm{g,max}}}\}\\
 0&\leq V_i\leq V_{i}^{\rm{max}}, \forall i
 \end{aligned}
\end{equation}
the energy storage capacity constraints of IC $i$ are always satisfied. 
\begin{proof}
{The mathematical induction is adopted to prove that the capacity constraints of $B_{i}(t)$ are always satisfied, $\forall i$. Clearly, the proposition holds at time slot 0. Then, the proposition is assumed to hold at time slot $t$, we prove that the proposition holds at time slot $t+1$ through the following four situations.}% and prove they also hold at time slot $t+1$.
\begin{enumerate}
\item Situation 1: $B_{i}(t)> \theta_{i}$. In this situation, $F_{i}(t) =B_{i}(t)-\theta_{i}>0$. According to the optimization problem (\ref{eq302}), $C_{i}^{\rm{e}}(t)=0$. Thus, $B_{i}(t+1) \leq B_{i}(t) \leq B_{i}^{\rm{max}}$.
\item Situation 2: $B_{i}(t)\leq\theta_{i}$. In this situation, since $C_{i}^{\rm{e}}(t) \leq{C_{i}^{\rm{e,max}}}$ and $\theta_{i}=V_ip^{\rm{e,max}}+D_{i}^{\rm{e,max}} \leq B_{i}^{\rm{max}}-C_{i}^{\rm{e,max}}$, $B_{i}(t+1) \leq B_{i}(t)+C_{i}^{\rm{e,max}} < \theta_{i} + C_{i}^{\rm{e,max}} \leq B_{i}^{\rm{max}}$.
\item Situation 3: $B_{i}(t)<D_{i}^{\rm{e,max}}$. In this situation, $F_{i}(t)=B_{i}(t)-\theta_{i}<D_{i}^{\rm{e,max}}-\theta_{i}=-V_ip^{\rm{e,max}}$. From (\ref{11a}), we can get $C_{i}^{\rm{e}}(t)-D_{i}^{\rm{e}}(t)=E_{i}(t)-E_{i}^{\rm{o}}(t)+X_{i}^{\rm{eb}}(t)-X_{i}^{\rm{es}}(t)+E_{i}^{\rm{CHP}}(t)-E_{i}^{\rm{p2g}}(t)+E_{i}^{\rm{r}}(t)-E_{i}^{\rm{d}}(t)$. According to (\ref{eq302}), we can get
\begin{equation}
\begin{aligned}
&\min_{\boldsymbol{M}_{i}(t)}  \Gamma_{i}(t)%\\&=\min_{\boldsymbol{M}_{i}(t)}  F_{i}(t)(C_{i}^{\rm{e}}(t)-D_{i}^{\rm{e}}(t))
%+Z_{i}(t)(C_{i}^{\rm{h}}(t)\\&-D_{i}^{\rm{h}}(t))
%+V_{i}(C_{i}^{\rm{c}}(t)+E_{i}(t)p^{\rm{e}}(t)-E_{i}^{\rm{o}}(t)p^{\rm{o}}(t)\\&+C_{i}^{\rm{X}}(t)-R_{i}^{\rm{S}}(t)+(G_{i}^{\rm{CHP}}(t)+G_{i}^{\rm{b}}(t))p^{\rm{g}}(t))
\\&=\min_{\boldsymbol{M}_{i}(t)}  (F_{i}(t)+V_{i}p^{\rm{e}}(t))E_{i}(t)+F_{i}(t)(-E_{i}^{\rm{o}}(t)\\&+X_{i}^{\rm{eb}}(t)-X_{i}^{\rm{es}}(t)+E_{i}^{\rm{CHP}}(t)-E_{i}^{\rm{p2g}}(t)\\&+E_{i}^{\rm{r}}(t)-E_{i}^{\rm{d}}(t))+Z_{i}(t)(C_{i}^{\rm{h}}(t)-D_{i}^{\rm{h}}(t))
\\&+V_{i}(C_{i}^{\rm{c}}(t)-E_{i}^{\rm{o}}(t)p^{\rm{o}}(t)+C_{i}^{\rm{X}}(t)-R_{i}^{\rm{S}}(t)\\&+(G_{i}^{\rm{CHP}}(t)+G_{i}^{\rm{b}}(t))p^{\rm{g}}(t))
\end{aligned}
\label{p1c3}
\end{equation}

Since $F_{i}(t)+V_{i}p^{\rm{e}}(t)<F_{i}(t)+V_ip^{\rm{e,max}}<0$, IC $i$ tends to increase $E_{i}(t)$, and $D_{i}^{\rm{e}}(t)=0$. Thus, $B_{i}(t+1) \geq B_{i}(t) \geq 0$.
\item Situation 4: $B_{i}(t)\geq D_{i}^{\rm{e,max}}$. In this situation, $B_{i}(t+1)= B_{i}(t)+C_{i}^{\rm{e}}(t)-D_{i}^{\rm{e}}(t)\geq B_{i}(t)-D_{i}^{\rm{e}}(t)\geq  0$.

\end{enumerate}

Combining the above four situations, Lemma 1 is still valid at time $t+1$, and it is proved.
Similarly, the capacity constraint of $W_{i}(t)$ can be guaranteed.
\end{proof}

Accordingly, the performance of our proposed method is given

\textbf{Theorem 1.} The time average total cost implemented by our proposed method satisfies:
\begin{align}
\lim_{T\rightarrow\infty}\frac{1}{T}\sum_{t=0}^{T-1}\mathbb{E}\{C_{i}(t)\}\leq C_{i}^{\rm{opt}}+\frac{A_i}{V_i}, \forall i
\end{align}
where $C_{i}^{\rm{opt}}$ denotes the minimum cost of the problem (\ref{eq301}). 

\begin{proof}
{$\forall i$, comparing this optimal solution (*) of problem} (\ref{eq302}) {got by minimizing $\Delta_{i}(t)+V_{i}\mathbb{E}\{C_{i}(t)\}$ with the result of the stationary random policy ($\Pi$), $\Delta_{i}(t)+V_{i}\mathbb{E}\{C_{i}(t)\}$ satisfies } %of relaxed problem (we use superscript $\Pi$).

\begin{equation}
\begin{aligned}
&\Delta_{i}(t)+V_{i}\mathbb{E}\{C_{i}(t)\}
 \\ & \leq A_{i}+\mathbb{E}\{F_{i}(t)(C_{i}^{\rm{e},*}(t)-D_{i}^{\rm{e},*}(t))+Z_{i}(t)(C_{i}^{\rm{h},*}(t)\\&-D_{i}^{\rm{h},*}(t))
 %\\&-V_{i}((\sum_{j_{i}=1}^{J_{i1}^e}(\frac{a_{j_{i}}^e}{p_{ie}^{*}(t)}-\frac{1}{tr_{j_{i}}^e})+\sum_{j_{i}=1}^{J_{i2}^e}E_{j_{i},min})p_{ie}^{*}(t)
%\\&+(\sum_{j_{i}=1}^{J_{i1}^h}(\frac{a_{j_{i}}^h}{p_{ih}^{*}(t)}-\frac{1}{tr_{j_{i}}^h})+\sum_{j_{i}=1}^{J_{i2}^h}H_{j_{i},min})p_{ih}^{*}(t)
+V_{i}(%L_{i}^{e,*}(t)p_{ie}^{*}(t)+L_{i}^{h,*}(t)p_{ih}^{*}(t)
%-a_{ie}L_{ie}^{*2}(t)+b_{ie}L_{ie}^{*}(t)-a_{ih}L_{ih}^{*2}(t)+b_{ih}L_{ih}^{*}(t)\\&
C_{i}^{\rm{c},*}(t)+E_{i}^{*}(t)p^{\rm{e}}(t)-E_{i}^{\rm{o},*}(t)p^{\rm{o}}(t)\\&+C_{i}^{\rm{X},*}(t)-R_{i}^{\rm{S},*}(t)+(G_{i}^{\rm{CHP},*}(t)+G_{i}^{\rm{b},*}(t))p^{\rm{g}}(t))\}
 \\ & \leq A_{i}+\mathbb{E}\{F_{i}(t)(C_{i}^{\rm{e},\Pi}(t)-D_{i}^{\rm{e},\Pi}(t))+Z_{i}(t)(C_{i}^{\rm{h},\Pi}(t)\\&-D_{i}^{\rm{h},\Pi}(t))
 %\\&-V_{i}((\sum_{j_{i}=1}^{J_{i1}^e}(\frac{a_{j_{i}}^e}{p_{ie}^{\Pi}(t)}-\frac{1}{tr_{j_{i}}^e})+\sum_{j_{i}=1}^{J_{i2}^e}E_{j_{i},min})p_{ie}^{\Pi}(t)
%\\&+(\sum_{j_{i}=1}^{J_{i1}^h}(\frac{a_{j_{i}}^h}{p_{ih}^{\Pi}(t)}-\frac{1}{tr_{j_{i}}^h})+\sum_{j_{i}=1}^{J_{i2}^h}H_{j_{i},min})p_{ih}^{\Pi}(t)
+V_{i}(%L_{i}^{e,\Pi}(t)p_{ie}^{\Pi}(t)+L_{i}^{h,\Pi}(t)p_{ih}^{\Pi}(t)
%-a_{ie}L_{ie}^{\Pi2}(t)+b_{ie}L_{ie}^{\Pi}(t)-a_{ih}L_{ih}^{\Pi2}(t)+b_{ih}L_{ih}^{\Pi}(t)\\&
C_{i}^{\rm{c},\Pi}(t)+E_{i}^{\Pi}(t)p^{\rm{e}}(t)-E_{i}^{\rm{o},\Pi}(t)p^{\rm{o}}(t)\\&+C_{i}^{\rm{X},\Pi}(t)-R_{i}^{\rm{S},\Pi}(t)+(G_{i}^{\rm{CHP},\Pi}(t)+G_{i}^{\rm{b},\Pi}(t))p^{\rm{g}}(t))\}%\sum_{k=1}^{n}(\alpha_k(t)+\Delta p_{i,k})X^{\Pi}_{i,k}(t) +\sum_{k=1}^{n}\alpha_i(t)S^{\Pi}_{k,i}(t))\}
%\leq & B_{i}+(Y_{i}(t)C^{*}_{i}(t)-V_{i}\hat{\alpha}_{i}(t)S^{*}_{i}(t))
%\\&+(V_{i}\hat{\beta}_{i}(t)X^{*}_{i}(t) - Y_{i}(t)D^{*}_{i}(t)
%\\&+V_{i}p_{e}(t)E^{*}_{i}(t)+V_{i}p_{g}P^{CHP,*}_{i}(t))
%\\&+H_{i}^{CHP,*}(t)(\eta_{hg}Z_{i}(t)+V_{i}p_{g})
%\\&+H^{b,*}_{i}(t)(\eta_{bg}Z_{i}(t)+V_{i}p_{g})
%\\&-Z_{i}(t)L_{i}^{w}(t)
%\\ \leq & B_{i}+(Y_{i}(t)C^{\Pi}_{i}(t)-V_{i}\hat{\alpha}_{i}(t)S^{\Pi}_{i}(t))
%\\&+(V_{i}\hat{\beta}_{i}(t)X^{\Pi}_{i}(t) - Y_{i}(t)D^{\Pi}_{i}(t)
%\\&+V_{i}p_{e}(t)E^{\Pi}_{i}(t)+V_{i}p_{g}P^{CHP,\Pi}_{i}(t))
%\\&+H_{i}^{CHP,\Pi}(t)(\eta_{hg}Z_{i}(t)+V_{i}p_{g})
%\\&+H^{b,\Pi}_{i}(t)(\eta_{bg}Z_{i}(t)+V_{i}p_{g})
%\\&-Z_{i}(t)L_{i}^{w}(t)
%\label{chap3equ:righthand}
\end{aligned}
\end{equation}

Based on (\ref{cd2}) and the policy ($\Pi$), $\Delta_{i}(t)+V_{i}\mathbb{E}\{C_{i}(t)\}$ satisfies
%According to Caratheodory Theory \cite{georgiadis2006resource}, 
%the optimal cost under random stationary policy is $C_{ir}^{opt}$%%The time average cost minimization of relaxed problem is of the relaxed problem is $C_{ir}^{opt}$
%, and according to (\ref{cd2}), we have:
%
\begin{equation}
\Delta_{i}(t)+V_{i}\mathbb{E}\{C_{i}(t)\} \leq A_{i} + V_{i}C_{ir}^{\rm{opt}} \leq A_{i} + V_{i}C_{i}^{\rm{opt}}
\end{equation}

Summing over $t \in \{0,1,\cdots,T-1\}$, the sum satisfies 
\begin{equation}
\mathbb{E}\{Y_{i}(T-1)-Y_{i}(0)\}+\sum_{t=0}^{T-1}V_{i}\mathbb{E}\{C_{i}(t)\} \leq TA_{i} + TV_{i}C_{i}^{\rm{opt}}
\end{equation}

Dividing both sides by $TV_{i}$ and taking $T \rightarrow \infty$, the time average cost satisfies 
\begin{equation}
\lim_{T \rightarrow \infty} \frac{1}{T} \sum_{t=0}^{T-1} \mathbb{E}\{ C_{i}(t) \} \leq C_{i}^{\rm{opt}} + \frac{A_{i}}{V_{i}}
\end{equation}

\end{proof}

According to Lemma 1 and Theorem 1, the cost implemented by our proposed method will arbitrarily approach the optimal cost as the capacities of the energy storage of IC $i$ increase.

\section{Simulation Result}

\subsection{Scene and Setup}
This paper considers the carbon and energy trading among four typical ICs in the Hongdou industrial park in Wuxi, Jiangsu Province, China. These four ICs are in turn: a biomedical industry, a plastics industry, an emerging research and development industry and a textile industry.  %That is, two process industries: biomedical industry (BI) and plastics industry (PI), an emerging research and development industry (ERDI) and a discrete manufacturing industry: textile industry (TI).
% The BI needs a large amount of process steam to maintain the production activities. The PI and TI needs much electricity to drive the wire drawing and weaving machines. The ERDI includes the research and development rooms, which consume large electrical loads for cooling. The ERDI provides technical support for the BI, PI and TI to improve the operation efficiency and production process, and prompt the renewal and upgrade of energy-intensive and high-emission equipment. 
 %Each IC has a battery, a PV system, a P2G device, a water tank, a CHP unit and a boiler. 
 The carbon emission intensities for gas-fired and coal-fired generation are 0.3 tCO$_2$/MWh and 0.85 tCO$_2$/MWh, respectively \cite{Cheng2019Modeling}.
Other relevant parameters are shown in Table I, which are similar to \cite{Zhu2022Stochastic}. 

\begin{table}
\centering
\caption{Parameters}\small
\begin{tabular}{m{2cm}m{1.5cm}|m{2cm}m{1.5cm}}
\hline
Parameter &Value & Parameter & Value \\
\hline
$B_{i}^{\rm{max}}$,$W_{i}^{\rm{max}}$& 4 MWh  & $p^{\rm{g}}$ &0.4 \textyen/KWh\\
$C_{i}^{\rm{h,max}}$,$D_{i}^{\rm{h,max}}$ & 0.4 MWh&$\eta_{i}^{\rm{p2g}}$& 60\%   \\
$C_{i}^{\rm{e,max}}$,$D_{i}^{\rm{e,max}}$& 0.4 MWh&$l$& 5 tCO$_2$ \\
%$\eta_{ipg}$&35\%  & $\eta_{ihg}$& 45\%\\
%$\eta_{cie}$,$\eta_{die}$,$\eta_{cih}$,$\eta_{dih}$&98\%& $\eta_{ibg}$&85\%  \\
$\alpha$,$\beta$&1& $\eta_{i}^{\rm{bg}}$&85\%  \\
\hline
\end{tabular}
\label{TAB1}
\end{table}

%To verify the performance of the proposed algorithm, the following different cases are introduced: 

%Case 1: Traditional energy scheduling with multi-energy conversion in an IC but without carbon and energy trading. 

%Case 2: Proposed two-stage energy trading and scheduling method but without carbon trading. 

%Case 3: Proposed two-stage carbon trading and energy scheduling method but without energy trading. 

\begin{figure}
\centering
\begin{minipage}{\linewidth}
  \centerline{\includegraphics[width=\hsize]{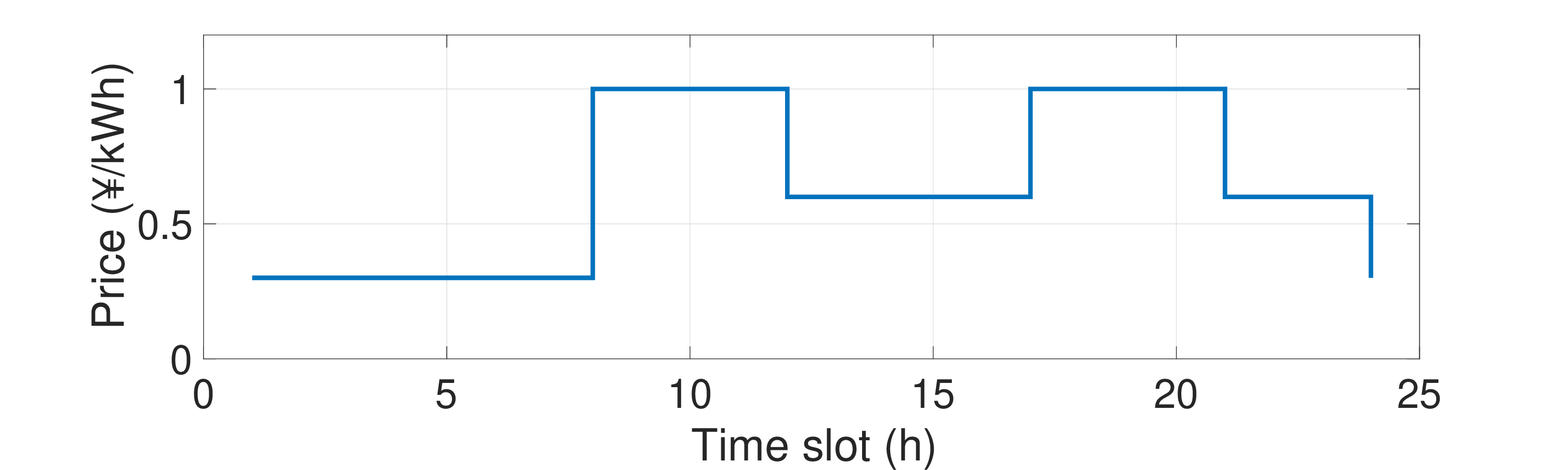}}
  \centerline{\scriptsize{(a) Price}}
\end{minipage}
%\hspace{-5pt}
\begin{minipage}{\linewidth}
  \centerline{\includegraphics[width=\hsize]{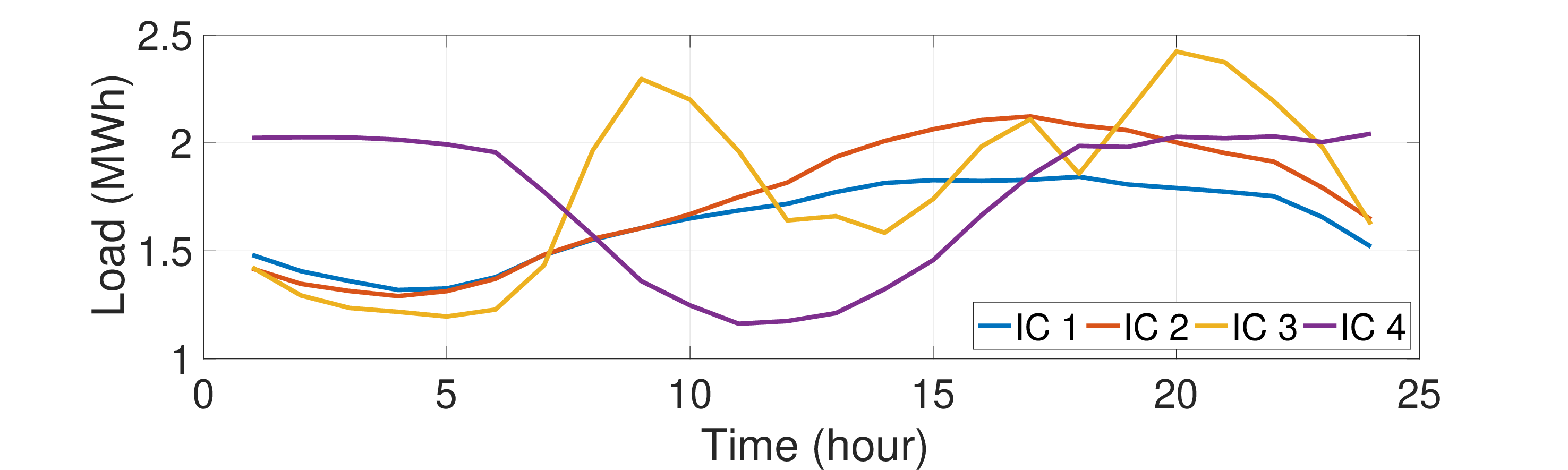}}
  \centerline{\scriptsize{(b) Electricity demand}}
\end{minipage}
\caption{Data of the industrial park.}
\label{fig5}
\end{figure}

To verify the proposed algorithm, the following two methods are introduced. Method 1 is a stochastic gradient-based algorithm (denoted as SGA) without renewable energy, which is similar to \cite{Deng2014Load}. 
%Case 2 is a stochastic optimization algorithm (denoted as OA) without energy storage, which is similar to \cite{Jia2011Multi}. 
{To evaluate the performance of Lyapunov optimization, Method 2 is adopted, where Lyapunov optimization is replaced by a game-based decision algorithm (denoted as GDA) without queue, similar to }\cite{Wang2014A}. {In addition, for fairness, the carbon and energy trading in the baseline methods are similar to our proposed method. The timescales for carbon and energy trading are one day and one hour, respectively.} Fig.~\ref{fig5} (a) shows the energy price of JiangSu Electric Power Company \cite{Illinois}.  
Fig.~\ref{fig5} (b) shows the load of the Hongdou industrial park \cite{Wang2021Multi}. 

%\begin{figure}
%\centering
%\begin{minipage}{\linewidth}
%  \centerline{\includegraphics[width=0.495\hsize]{price.eps}}
%  \centerline{\scriptsize{(a) }}
%\end{minipage}
%\begin{minipage}{\linewidth}
%  \centerline{\includegraphics[width=0.495\hsize]{elb.eps}}
%  \centerline{\scriptsize{(b) }}
%\end{minipage}
%\caption{Data of the industrial park. (a) Electricity price of the power company. (b) Electricity demand of four ICs. }
%\label{fig5}
%\end{figure} and improve the energy structure

\subsection{Results}
\begin{figure}
  \centering
  \includegraphics[width=\hsize]{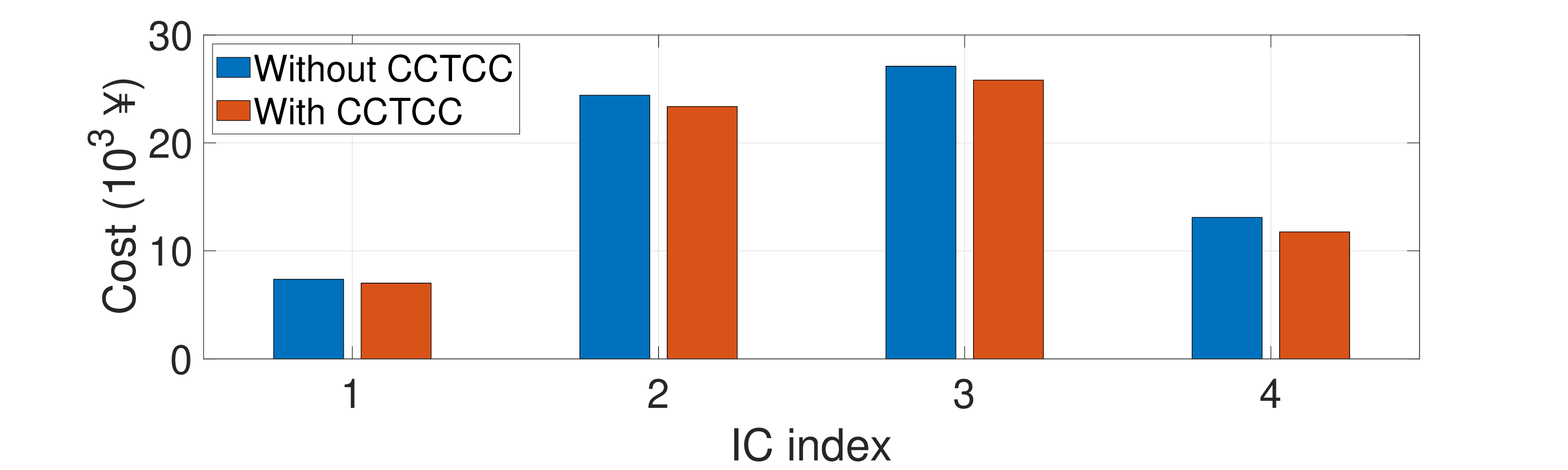}
  \caption{{Total costs of four ICs with and without CCTCC.}}
  \label{fig6}
\end{figure}

Coordinating carbon trading and carbon capture (CCTCC) helps high-emitting ICs comply with emission limits.  The total costs of the ICs are lower with CCTCC, as shown in Fig.~\ref{fig6}. The total costs of ICs consist of energy cost from the CCPP, the gas plant and other ICs, carbon trading cost and green certificate income, where the timescale of carbon trading is set as one day. {The calculation results with and without CCTCC for ICs are denoted as Tables II and III. The ICs without CCTCC need to buy electricity from the CCPP with maximum carbon capture power (CCPPMP). Otherwise, they will pay huge fines. The cost of carbon capture is transferred to the ICs, which incurs the prices of CCPPMP to be higher than grid electricity prices. By using the electricity from the CCPPMP, the ICs with additional carbon rights  can earn green certificate revenue. For IC $i$ with CCTCC, when the transaction cost exceeds the carbon capture cost, the IC will capture this part of carbon emissions through carbon capture. The remaining carbon emissions are transferred to ICs with lower carbon emissions in the form of carbon trading to achieve a balance between carbon emissions and carbon quotas.} The efficiency of CCPPMP is 85\%, and the carbon trading price $p^{\rm{c}}=100$ \textyen/tCO$_2$. %The reward and punishment factors are $\alpha=\beta=1$.}

% Through trading carbon, the costs of ICs 1 and 2 reduce more, and there is a slight cost reduction for IC 3. Since the carbon emission of IC 1 is much higher than the allocated carbon emission and IC 2 has little carbon emissions, IC 1 purchases the carbon emission right from IC 2 instead of paying huge fines for carbon emissions violations and IC 2 obtains extra benefits. The carbon emissions generated by IC 3 are roughly equivalent to the allocated carbon emissions. Therefore, the cost of ICs 1 and 2 is greatly reduced, and the cost of IC 3 does not change much.

\begin{table}\footnotesize
\setlength\tabcolsep{2pt}
\centering
\caption{{Calculation result without CCTCC for ICs in one day}}
\begin{tabular}{l l l l  l  }
\hline
IC& IC1& IC2& IC3& IC4   \\
From the CCPPMP (MWh)& 11.42  & 14.64& 16.55 & 16.58  \\
From the gas plant (MWh)& 19.62  & 26.39& 33.75&20.48  \\
Energy cost ($* 10^3$ \textyen) & 8.56  & 25.25& 27.83&14.22\\
%Energy cost ($* 10^3$ \textyen) & 9.19  & 25.17& 29.13\\
%Carbon emission (tCO$_2$)& 6.02  & 10.09& 11.33\\
Carbon emission (tCO$_2$)& 7.34  & 9.78& 12.24& 8.26\\
Carbon capture (tCO$_2$)& 8.25  & 10.58& 11.96&11.98\\
%Carbon capture (tCO$_2$)& 11.39*0.85^2  & 14.64*0.85^2& 16.41*0.85^2&16.61*0.85^2
Carbon quota (tCO$_2$)& 15.67  & 16.64& 17.10& 16.77\\
%Green certificate income ($* 10^3$ \textyen) & 0.72  & 0.41& 0.33\\
%Total cost ($* 10^3$ \textyen) & 8.47  & 25.76& 28.80\\
Green certificate revenue ($* 10^3$ \textyen) & 1.17  & 0.87& 0.49& 1.20\\
Total cost ($* 10^3$ \textyen) & 7.39  & 24.38& 27.34&13.02\\
\hline
\end{tabular}
%\end{table}
\vspace{1em}
%\begin{table}[H]
\setlength\tabcolsep{2pt}
\centering
\caption{{Calculation result with CCTCC for ICs in one day}}
\begin{tabular}{l l l l  l  }
\hline
IC & IC1& IC2& IC3& IC4   \\
%From CCPS (MWh)& 1.79  & 6.97& 8.29\\捕集耗电500度/吨，在电价为0.3元时，碳捕集成本为150元/吨，5吨以内的碳价为100元/吨，5吨以上的碳价为200元/吨。碳捕集率为85%
From the CCPP (MWh)& 13.41  & 17.48&18.66&19.47  \\
From the gas plant (MWh)& 17.52  & 24.83& 31.89 & 17.02\\
Energy cost ($* 10^3$ \textyen) & 6.91  & 22.86& 25.32& 11.26\\
%Energy cost ($* 10^3$ \textyen) & 6.85  & 23.03& 24.92\\
%Carbon emission (tCO$_2$)& 16.61  & 24.09& 26.74\\
%Carbon trading cost ($* 10^3$ \textyen) & 0.05  & 0.50& 0.71\\
%Total cost  ($* 10^3$ \textyen) & 6.90  & 23.53& 25.63\\
Carbon emission (tCO$_2$)& 16.65  & 22.31& 25.43& 21.66\\
Carbon capture (tCO$_2$)& 0  & 0.67& 3.33& 0\\
%碳捕集成本0&0.1&0.5&0，因此能量成本为6.91  & 22.76+0.1& 24.82+0.5& 11.26
Carbon quota (tCO$_2$)& 15.67  & 16.64& 17.10& 16.77\\
Carbon trading cost ($* 10^3$ \textyen) & 0.10  & 0.50& 0.50& 0.49\\
Total cost  ($* 10^3$ \textyen) & 7.01  & 23.36& 25.82& 11.75\\
\hline
\end{tabular}
\end{table}

\begin{figure}
\centering
\begin{minipage}{\linewidth}
  \centerline{\includegraphics[width=\hsize]{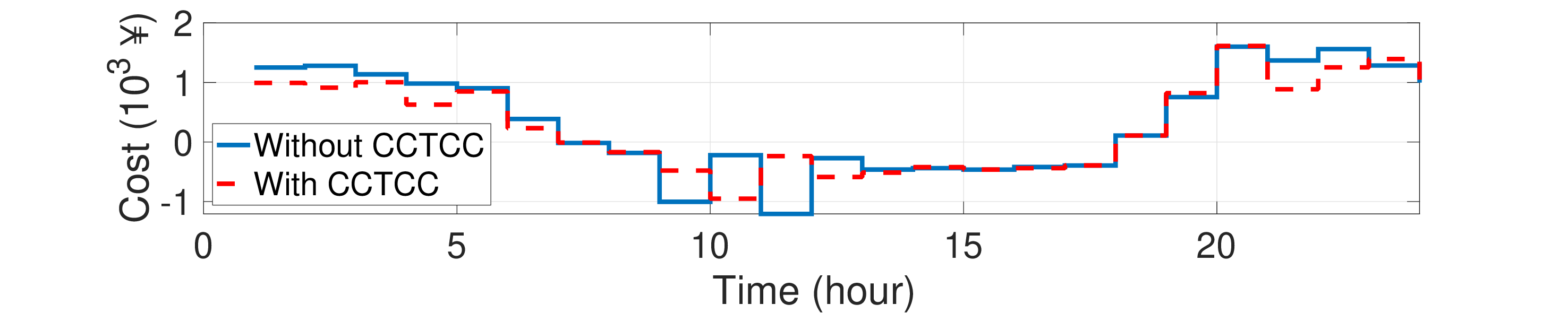}}
  \centerline{\scriptsize{(a) IC 1}}
\end{minipage}
\begin{minipage}{\linewidth}
  \centerline{\includegraphics[width=\hsize]{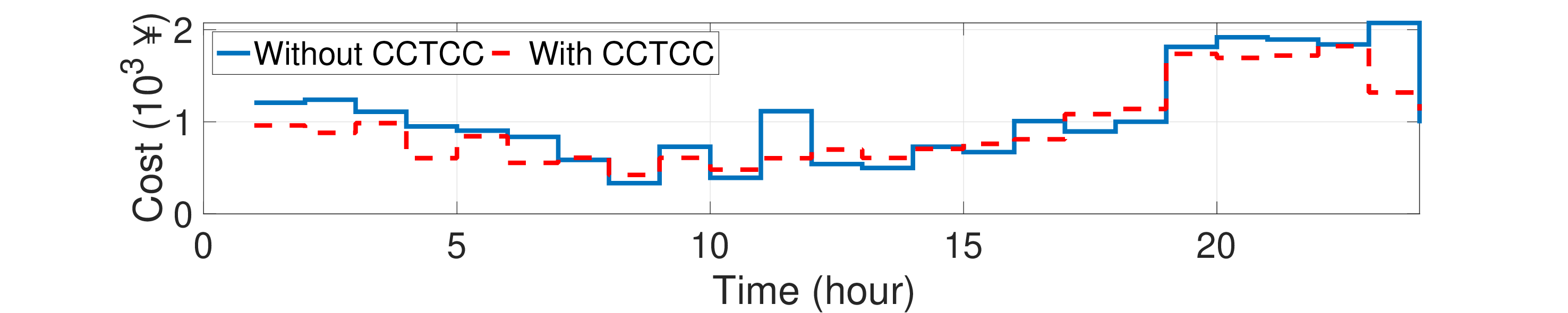}}
  \centerline{\scriptsize{(b) IC 2}}
\end{minipage}
\begin{minipage}{\linewidth}
  \centerline{\includegraphics[width=\hsize]{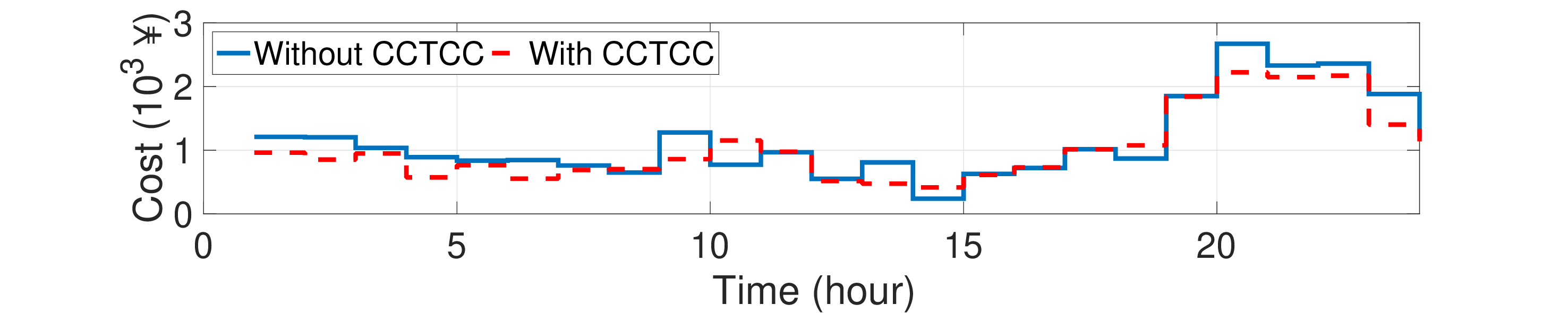}}
  \centerline{\scriptsize{(c) IC 3}}
\end{minipage}
\begin{minipage}{\linewidth}
  \centerline{\includegraphics[width=\hsize]{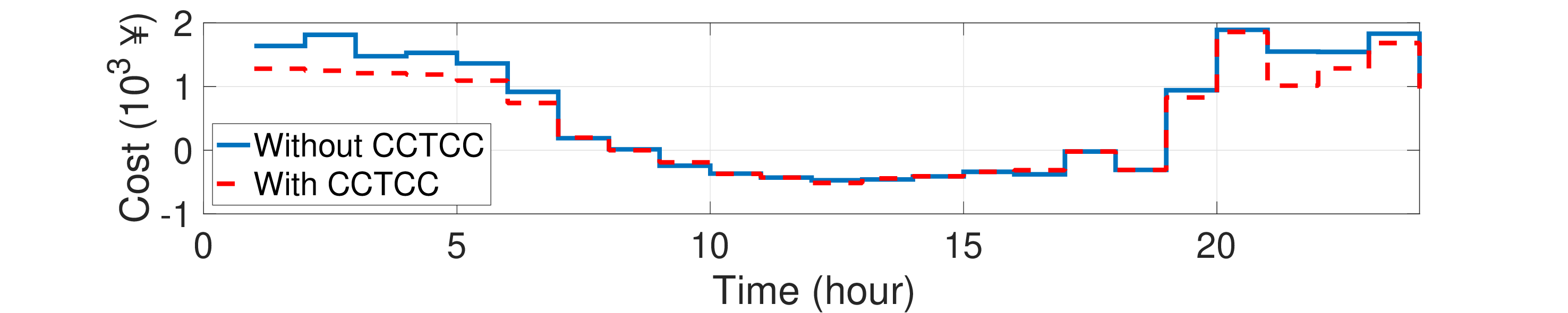}}
  \centerline{\scriptsize{(d) IC 4}}
\end{minipage}
\caption{Cost comparison for four ICs with and without CCTCC.}
\label{fig7c}
\end{figure}

\begin{figure}
\centering
\begin{minipage}{\linewidth}
  \centerline{\includegraphics[width=\hsize]{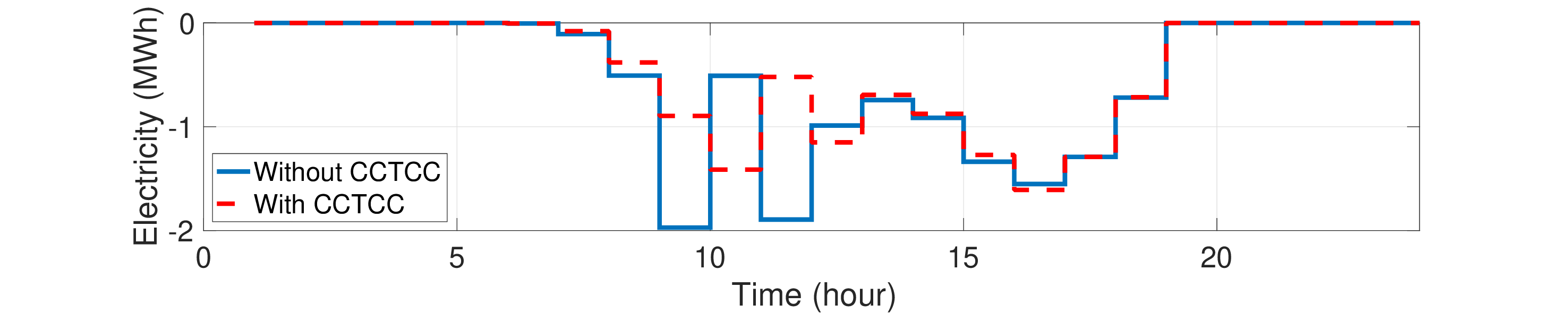}}
  \centerline{\scriptsize{(a) IC 1}}
\end{minipage}
\begin{minipage}{\linewidth}
  \centerline{\includegraphics[width=\hsize]{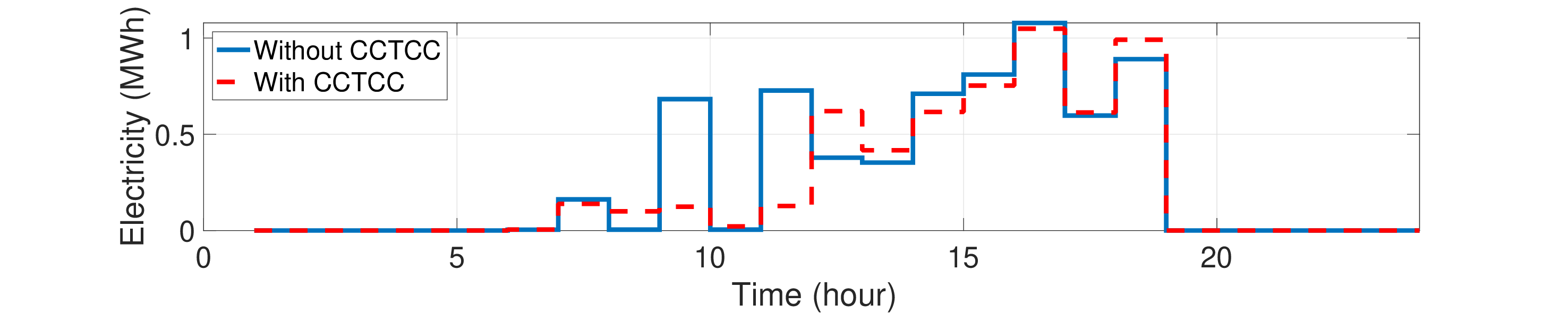}}
  \centerline{\scriptsize{(b) IC 2}}
\end{minipage}
\begin{minipage}{\linewidth}
  \centerline{\includegraphics[width=\hsize]{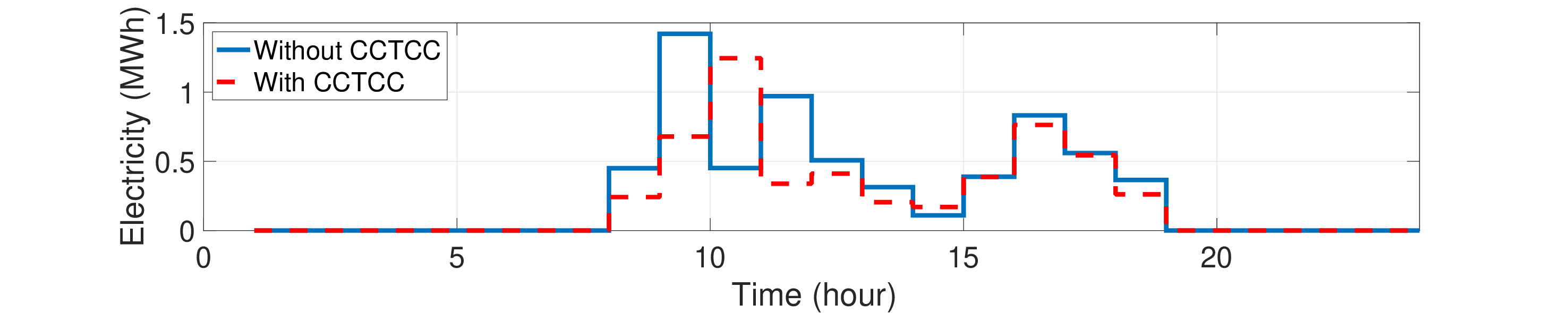}}
  \centerline{\scriptsize{(c) IC 3}}
\end{minipage}
\begin{minipage}{\linewidth}
  \centerline{\includegraphics[width=\hsize]{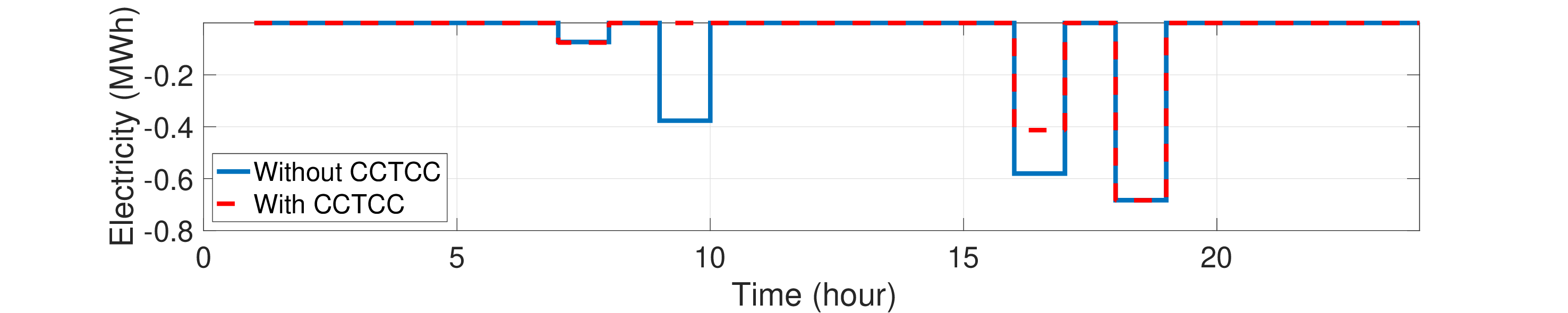}}
  \centerline{\scriptsize{(d) IC 4}}
\end{minipage}
\caption{Electricity trading for four ICs with and without CCTCC.}
\label{fig7e}
\end{figure}

\begin{figure}
\centering
\begin{minipage}{\linewidth}
  \centerline{\includegraphics[width=\hsize]{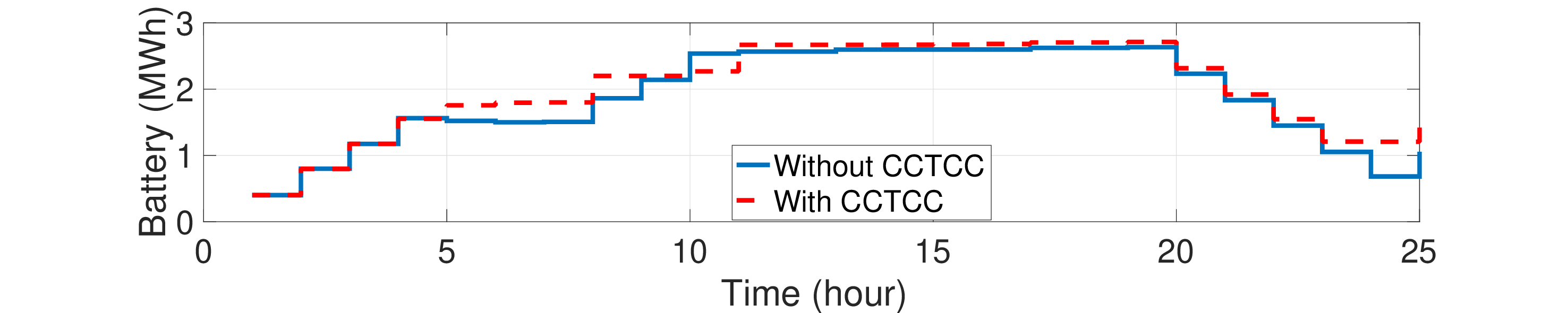}}
  \centerline{\scriptsize{(a) IC 1}}
\end{minipage}
\begin{minipage}{\linewidth}
  \centerline{\includegraphics[width=\hsize]{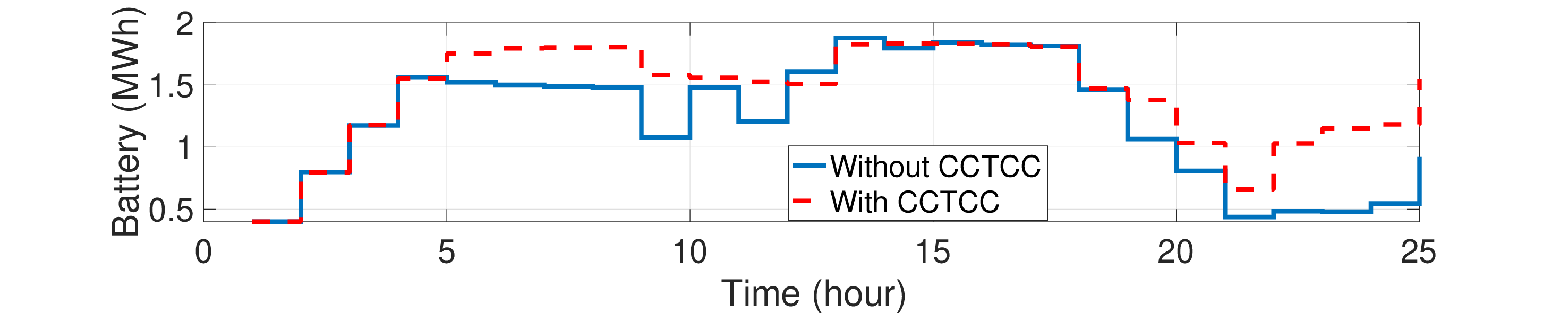}}
  \centerline{\scriptsize{(b) IC 2}}
\end{minipage}
\begin{minipage}{\linewidth}
  \centerline{\includegraphics[width=\hsize]{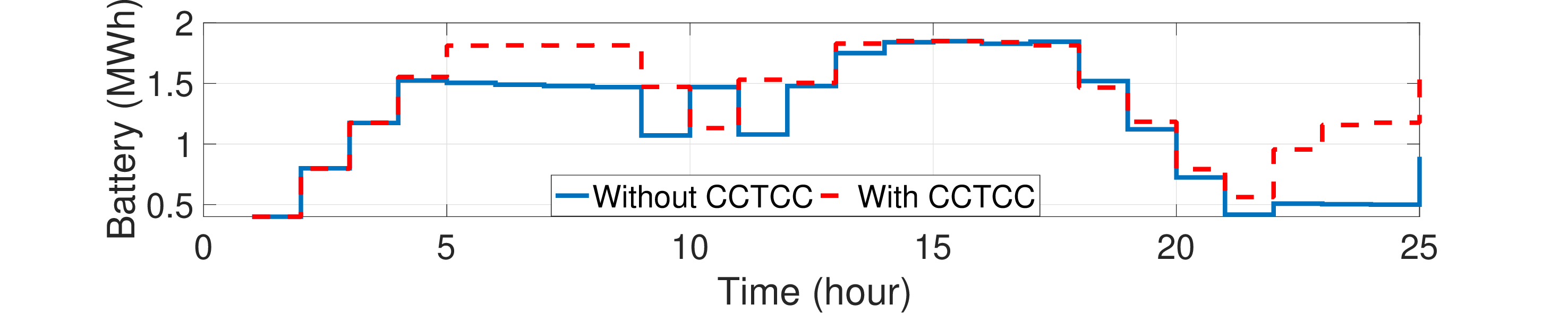}}
  \centerline{\scriptsize{(c) IC 3}}
\end{minipage}
\begin{minipage}{\linewidth}
  \centerline{\includegraphics[width=\hsize]{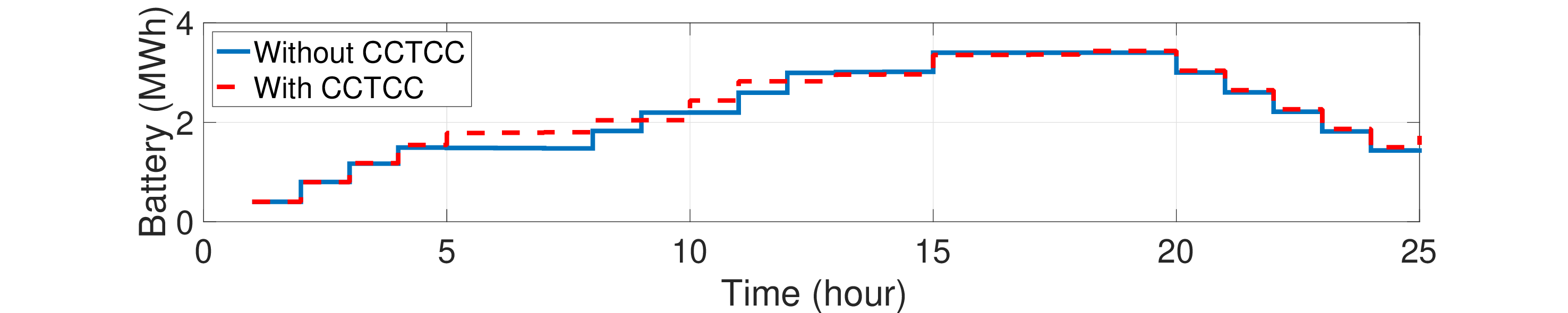}}
  \centerline{\scriptsize{(d) IC 4}}
\end{minipage}
\caption{Battery level for four ICs with and without CCTCC.}
\label{fig7b}
\end{figure}

The cost, electricity trading and battery dynamics for four ICs with and without CCTCC are shown in Fig.~\ref{fig7c}-\ref{fig7b}. In trading,  a negative value means selling energy, a positive value means purchasing energy. ICs have lower costs under CCTCC in most time slots, as shown in Fig.~\ref{fig7c}, where ICs with CCTCC can flexibly coordinate the amount of carbon capture and carbon trading based on their cost, whereas ICs without CCTCC may purchase high-priced electricity from CCPPMP or pay huge fines for carbon emissions violations. To reduce the amount of high price electricity purchased from CCPPMP, ICs 2 and 3 without CCTCC purchase more electricity from ICs 1 and 4 to reduce the energy cost, as shown in Fig.~\ref{fig7e}. For the same reason, ICs without CCTCC charge less electricity into batteries, which also reduces the purchase of the high-priced electricity from CCPPMP, as shown in Fig.~\ref{fig7b}. 

%\begin{figure*}
%\centering
%\begin{minipage}{\linewidth}
%{\includegraphics[width=0.33\hsize]{cct1.eps}}
%{\includegraphics[width=0.33\hsize]{ect1.eps}}
%%{\includegraphics[width=0.5\hsize]{gct1.eps}}
%{\includegraphics[width=0.33\hsize]{bct1.eps}}
%  \centerline{\scriptsize{(a) IC 1}}
%\end{minipage}
%\begin{minipage}{\linewidth}
%{\includegraphics[width=0.33\hsize]{cct2.eps}}
%{\includegraphics[width=0.33\hsize]{ect2.eps}}
%%\centerline{\includegraphics[width=\hsize]{gct2.eps}}
%{\includegraphics[width=0.33\hsize]{bct2.eps}}
%  \centerline{\scriptsize{(b) IC 2}}
%\end{minipage}
%\begin{minipage}{\linewidth}
%{\includegraphics[width=0.33\hsize]{cct3.eps}}
%{\includegraphics[width=0.33\hsize]{ect3.eps}}
%%\centerline{\includegraphics[width=\hsize]{gct3.eps}}
%{\includegraphics[width=0.33\hsize]{bct3.eps}}
%  \centerline{\scriptsize{(c) IC 3}}
%\end{minipage}
%\begin{minipage}{\linewidth}
%{\includegraphics[width=0.33\hsize]{cct4.eps}}
%{\includegraphics[width=0.33\hsize]{ect4.eps}}
%%\centerline{\includegraphics[width=\hsize]{gct3.eps}}
%{\includegraphics[width=0.33\hsize]{bct4.eps}}
%  \centerline{\scriptsize{(d) IC 4}}
%\end{minipage}
%\caption{Cost, electricity trading and battery dynamics comparison for four ICs with and without CCTCC.}
%\label{fig7}
%\end{figure*}

\begin{figure}
  \centering
  \includegraphics[width=\hsize]{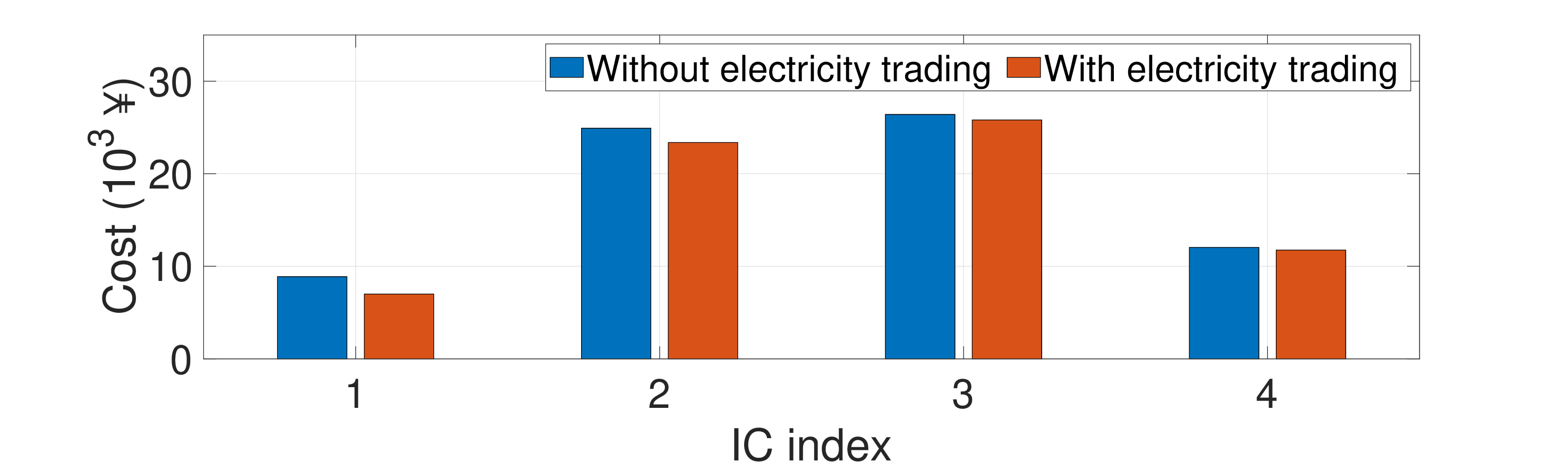}
  \caption{Total costs of four ICs with and without electricity trading.}
  \label{fig8}
\end{figure}

\begin{figure}
\centering
\begin{minipage}{\linewidth}
  \centerline{\includegraphics[width=\hsize]{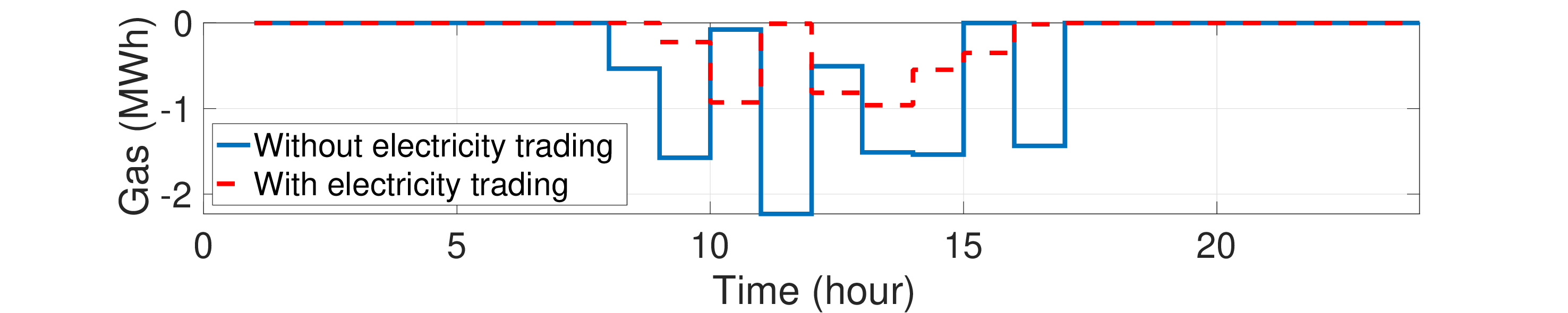}}
  \centerline{\scriptsize{(a) IC 1}}
\end{minipage}
\begin{minipage}{\linewidth}
  \centerline{\includegraphics[width=\hsize]{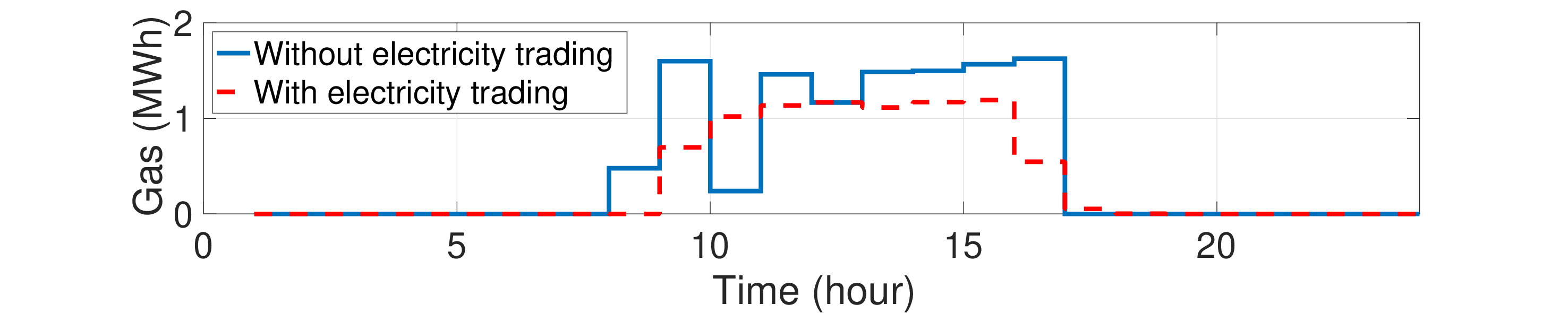}}
  \centerline{\scriptsize{(b) IC 2}}
\end{minipage}
\begin{minipage}{\linewidth}
  \centerline{\includegraphics[width=\hsize]{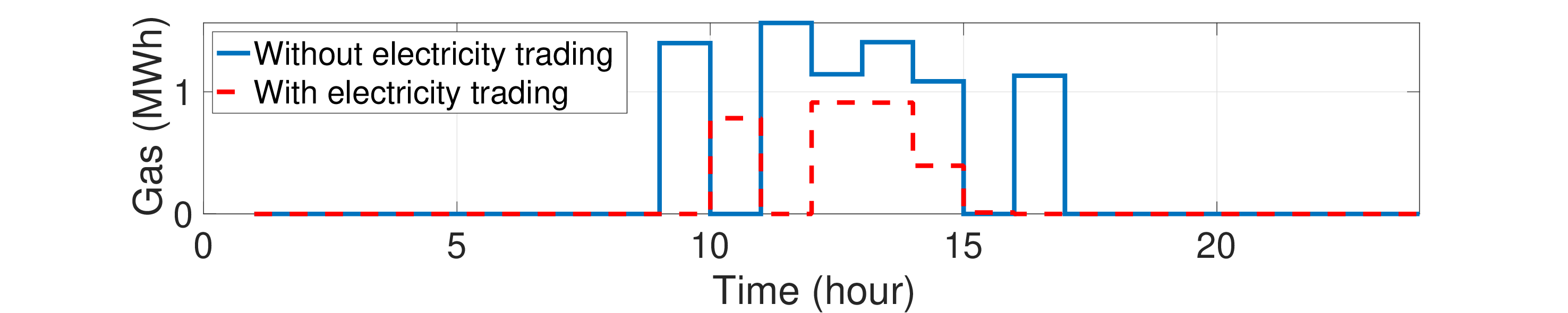}}
  \centerline{\scriptsize{(c) IC 3}}
\end{minipage}
\begin{minipage}{\linewidth}
  \centerline{\includegraphics[width=\hsize]{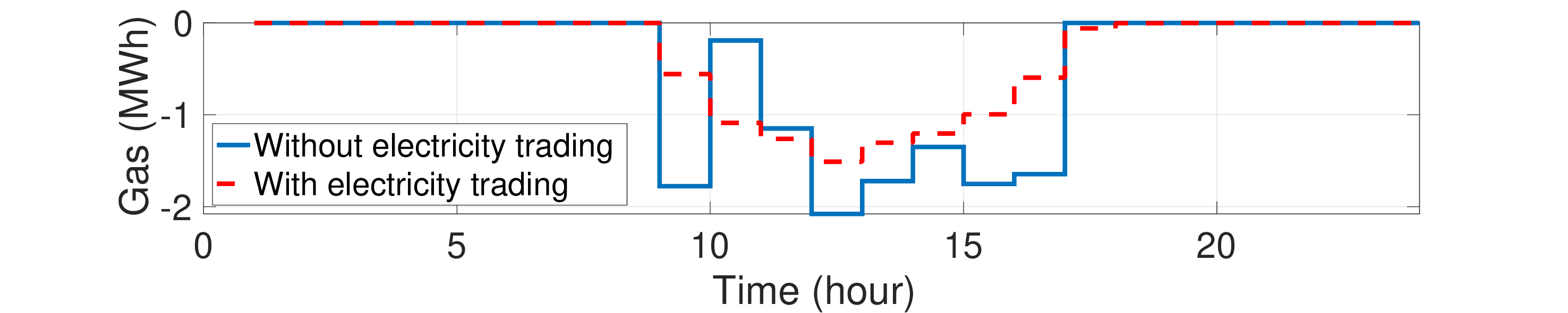}}
  \centerline{\scriptsize{(d) IC 4}}
\end{minipage}
\caption{Gas trading for four ICs with and without electricity trading.}
\label{fig9}
\end{figure}

Energy trading helps ICs alleviate energy supply and demand imbalance. Fig.~\ref{fig8} shows that with electricity trading, the cost of the four ICs are reduced. According to Fig.~\ref{fig9}, when there is no electricity trading among the ICs, the four ICs trade more gas to make up the energy deficiency. Thus, the cost reduction is not significant.
%Since IC 1 generates more electricity by its PV system, it can sell electricity to ICs 2 and 3 instead of selling to the power grid at a lower price. Thus, the cost of IC 1 is reduced more. IC 2 generates much electricity and heat by its CHP unit to make up the deficient electricity which cannot be purchased from other ICs. Then, IC 2 obtains some revenues by selling the extra heat to ICs 1 and 3. Thus, although IC 2 cannot purchase the low-price electricity from other ICs, it still does not incur high costs. 

%\begin{figure}
%\centering
%\begin{minipage}{\linewidth}
%  \centerline{\includegraphics[width=\hsize]{get1.eps}}
%  \centerline{\scriptsize{(a) IC 1}}
%\end{minipage}
%\begin{minipage}{\linewidth}
%  \centerline{\includegraphics[width=\hsize]{get2.eps}}
%  \centerline{\scriptsize{(b) IC 2}}
%\end{minipage}
%\begin{minipage}{\linewidth}
%  \centerline{\includegraphics[width=\hsize]{get3.eps}}
%  \centerline{\scriptsize{(c) IC 3}}
%\end{minipage}
%\begin{minipage}{\linewidth}
%  \centerline{\includegraphics[width=\hsize]{get4.eps}}
%  \centerline{\scriptsize{(c) IC 4}}
%\end{minipage}
%\caption{Gas trading comparison for four ICs.}
%\label{fig9}
%\end{figure}

%\begin{figure*}[htb]
%\centering
%{\includegraphics[width=0.245\hsize]{get1.eps}}
%{\includegraphics[width=0.245\hsize]{get2.eps}}
%{\includegraphics[width=0.245\hsize]{get3.eps}}
%{\includegraphics[width=0.245\hsize]{get4.eps}}
%\caption{Gas trading comparison for four ICs. (a) IC 1. (b) IC 2. (c) IC 3. (d) IC 4.}
%\label{fig9}
%\end{figure*}

%\begin{minipage}{\linewidth}
%{\includegraphics[width=0.33\hsize]{cct4.eps}}
%{\includegraphics[width=0.33\hsize]{ect4.eps}}
%{\includegraphics[width=0.33\hsize]{bct4.eps}}
%  \centerline{\scriptsize{(c) IC 4}}
%\end{minipage}
%\caption{Cost, electricity trading and battery dynamics comparison for four ICs.}
%\label{fig7}
%\end{figure*}

\begin{figure}[ht]
  \centering
  \includegraphics[width=\hsize]{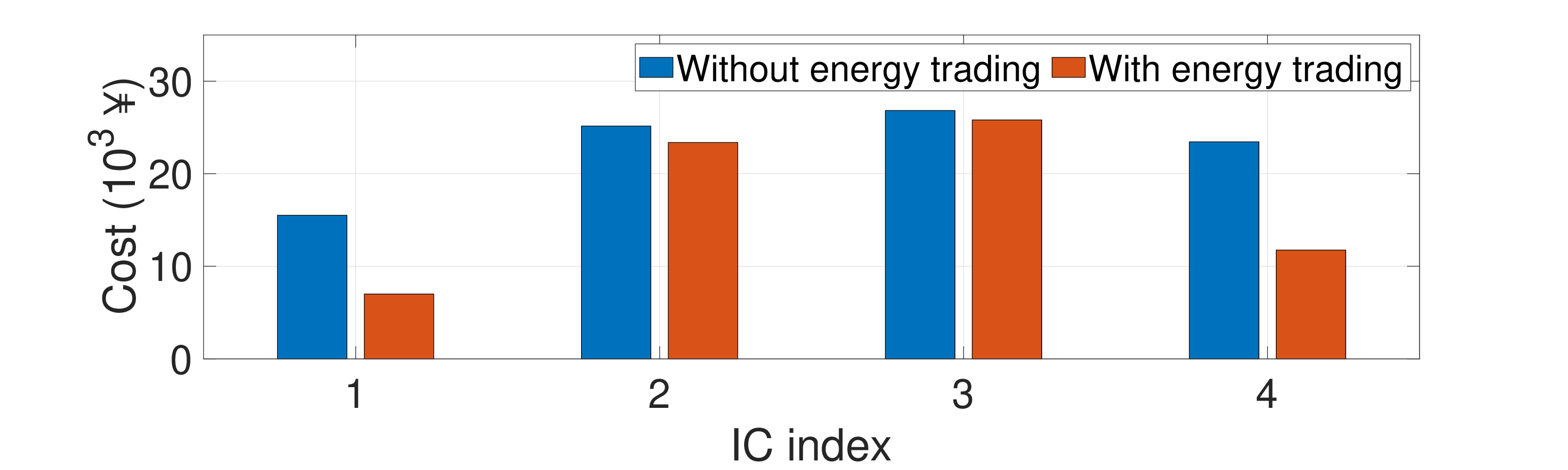}
  \caption{Total costs of four ICs with and without energy trading.}
  \label{fig10}
\end{figure}

\begin{figure}
\centering
\begin{minipage}{\linewidth}
  \centerline{\includegraphics[width=\hsize]{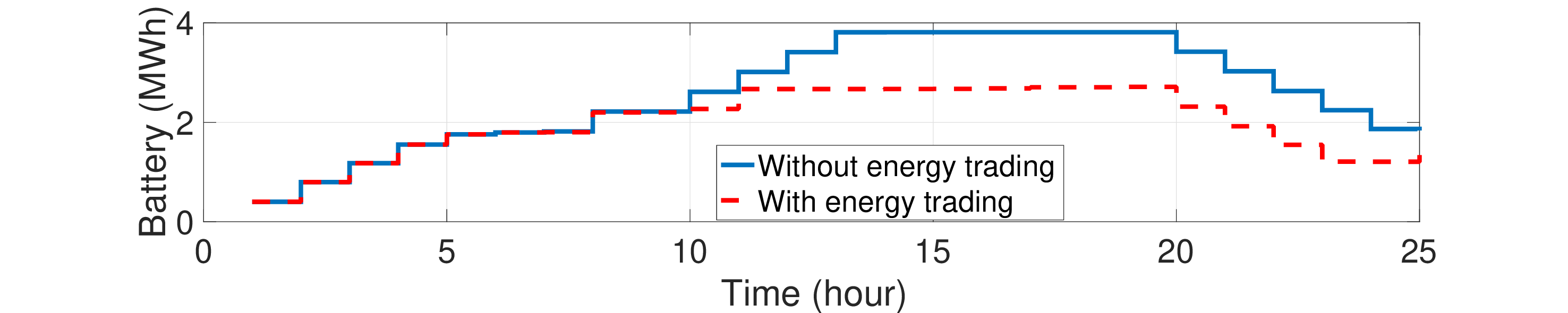}}
  \centerline{\scriptsize{(a) IC 1}}
\end{minipage}
%\hspace{-5pt}
\begin{minipage}{\linewidth}
  \centerline{\includegraphics[width=\hsize]{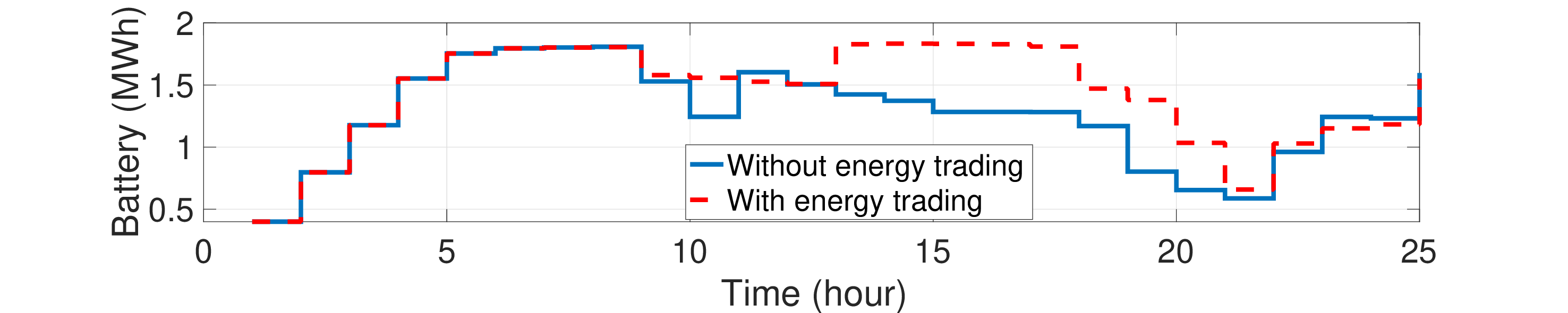}}
  \centerline{\scriptsize{(b) IC 2}}
\end{minipage}
\begin{minipage}{\linewidth}
  \centerline{\includegraphics[width=\hsize]{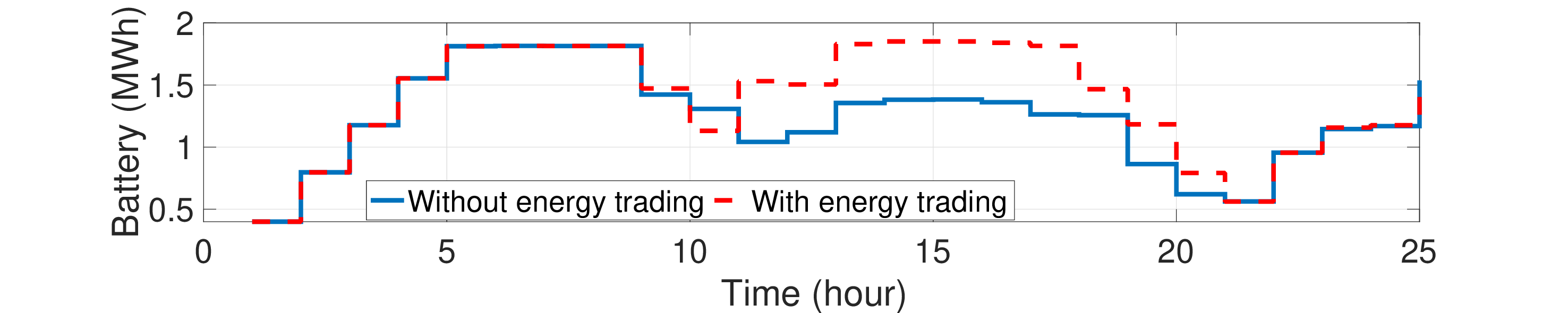}}
  \centerline{\scriptsize{(c) IC 3}}
\end{minipage}
\begin{minipage}{\linewidth}
  \centerline{\includegraphics[width=\hsize]{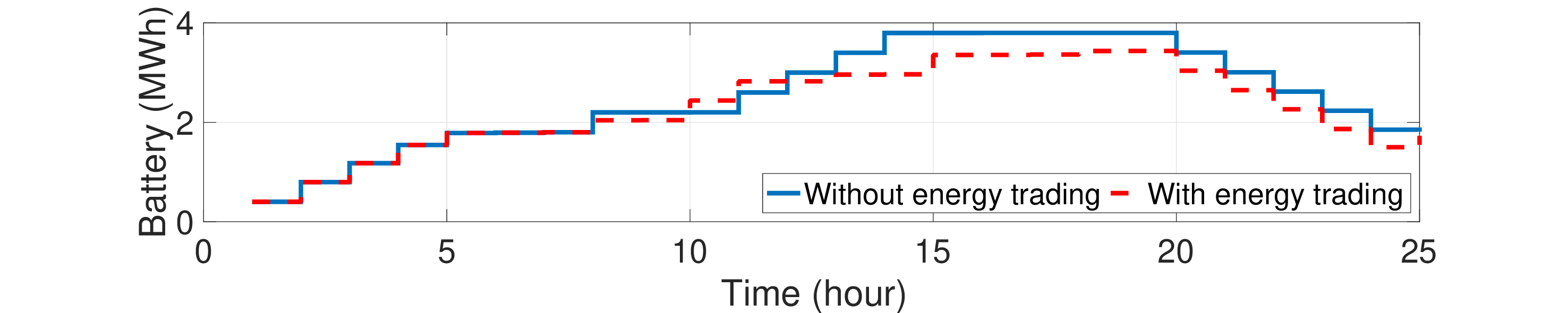}}
  \centerline{\scriptsize{(d) IC 4}}
\end{minipage}
\caption{Battery level for four ICs with and without energy trading.}
\label{fig11}
\end{figure}

Fig.~\ref{fig10} shows that the ICs reduce costs through electricity and gas trading, where there is a significant cost reduction for ICs 1 and 4, and a slight cost reduction for ICs 2 and 3. With electricity and gas trading, ICs 1 and 4 can sell electricity and gas to ICs 2 and 3, instead of selling electricity at low prices to the power grid or storing large amounts of energy. Fig.~\ref{fig11} shows that ICs 1 and 4 without energy trading charge more electricity. Thus, the costs of ICs 1 and 4 are greatly reduced.  Although ICs 2 and 3 with energy trading can purchase low-priced energy from ICs 1 and 4, ICs 2 and 3 without energy trading can generate electricity and heat by CHP instead of purchasing high-priced electricity from the power grid. Thus, the cost reduction for ICs 2 and 3 is less. Additionally, Fig.~\ref{fig11} shows that ICs 2 and 3 with energy trading charge more electricity during 12:00-17:00 and discharge electricity to supply the electricity demand during the high-priced periods 17:00-21:00, which reflects the advantages of energy storage. 

%\begin{figure}
%\centering
%\begin{minipage}{\linewidth}
%  \centerline{\includegraphics[width=\hsize]{bent1.eps}}
%  \centerline{\scriptsize{(a) IC 1}}
%\end{minipage}
%\begin{minipage}{\linewidth}
%  \centerline{\includegraphics[width=\hsize]{bent2.eps}}
%  \centerline{\scriptsize{(b) IC 2}}
%\end{minipage}
%\begin{minipage}{\linewidth}
%  \centerline{\includegraphics[width=\hsize]{bent3.eps}}
%  \centerline{\scriptsize{(c) IC 3}}
%\end{minipage}
%\begin{minipage}{\linewidth}
%  \centerline{\includegraphics[width=\hsize]{bent4.eps}}
%  \centerline{\scriptsize{(d) IC 4}}
%\end{minipage}
%\caption{Battery level comparison for four ICs.}
%\label{fig11}
%\end{figure}

%\begin{figure*}
%\centering
%{\includegraphics[width=0.245\hsize]{bent1.eps}}
%{\includegraphics[width=0.245\hsize]{bent2.eps}}
%{\includegraphics[width=0.245\hsize]{bent3.eps}}
%{\includegraphics[width=0.245\hsize]{bent4.eps}}
%\caption{Battery level comparison for four ICs. (a) IC 1. (b) IC 2. (c) IC 3. (d) IC 4.}
%\label{fig11}
%\end{figure*}, which is significant for sustainable development

\begin{figure}
  \centering
  \includegraphics[width=\hsize]{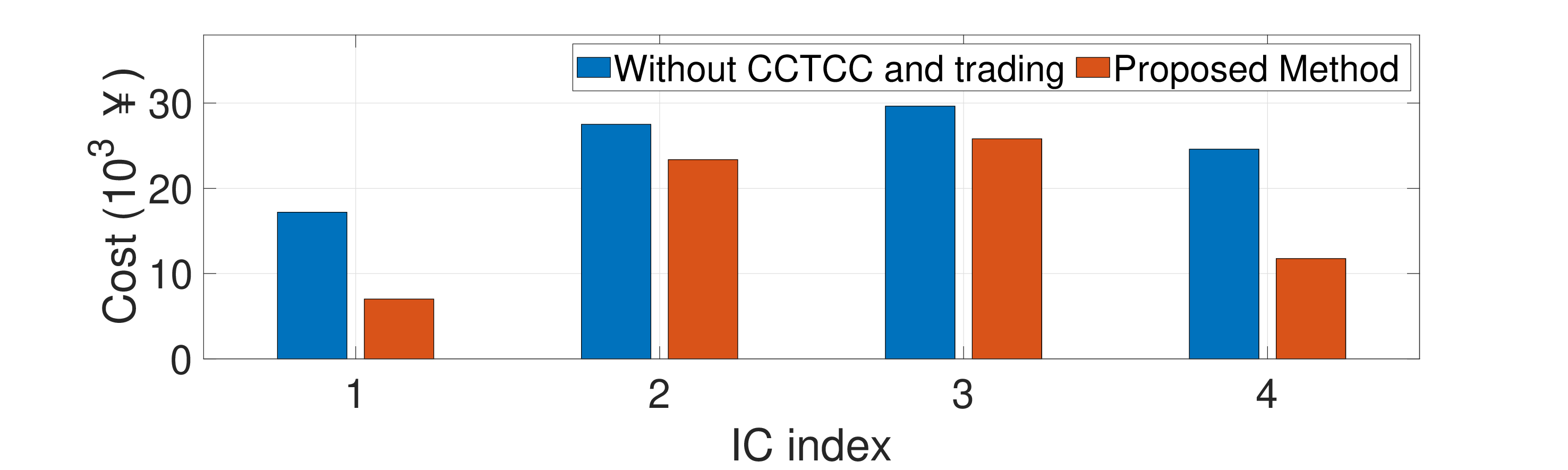}
  \caption{Total cost comparison of our proposed method with the method lacking CCTCC and energy trading.}
  \label{pmwt}
\end{figure}

Since carbon and energy trading are not popularized at present, the total cost comparison of our proposed method with the method lacking CCTCC and energy trading is given in Fig.~\ref{pmwt}, which demonstrates the effectiveness of the proposed method.

\begin{figure}[ht]
  \centering
  \includegraphics[width=0.75\hsize]{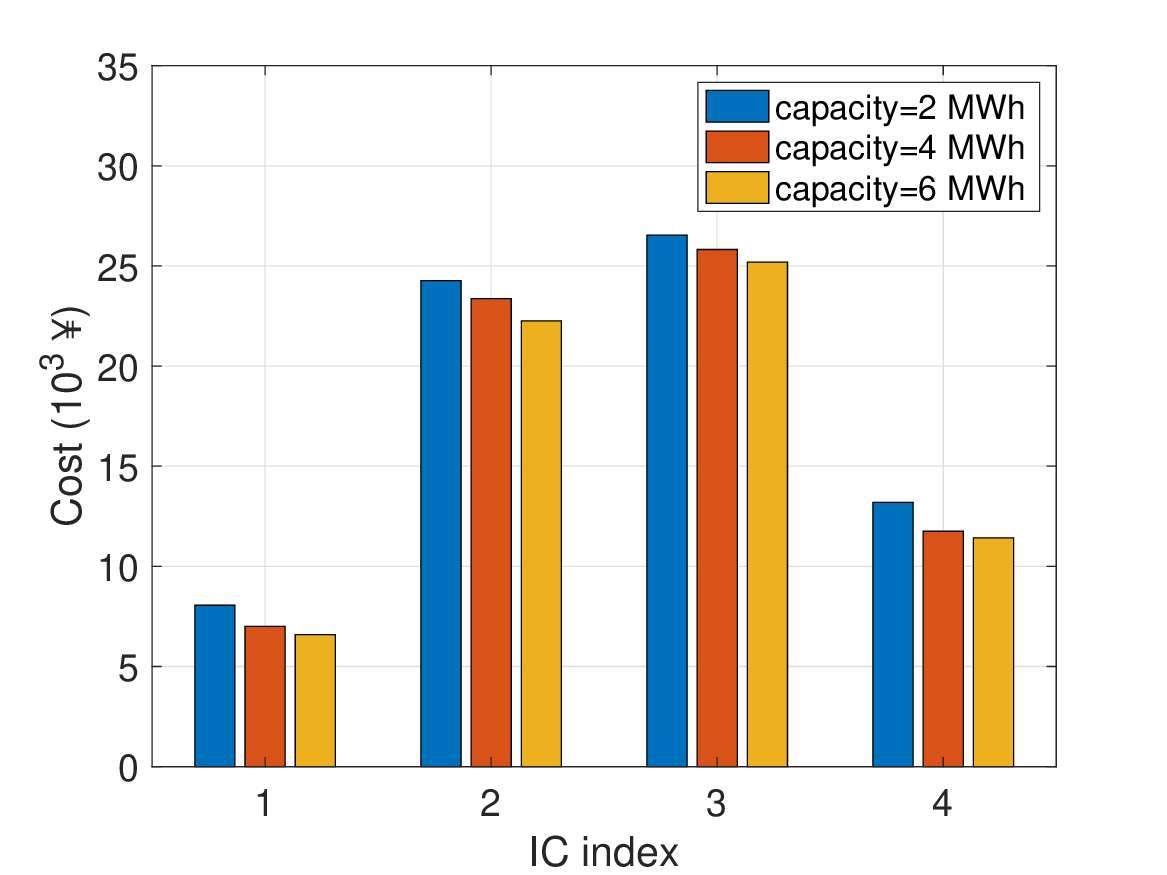}
  \caption{Total costs of four ICs under different energy storage capacities.}
  \label{escc}
\end{figure}

To confirm the performance of Lyapunov optimization, the total costs of four ICs under different energy storage (i.e., battery and water tank) capacities are shown in Fig.~\ref{escc}. The simulation shows that the larger the energy storage capacity, the lower the cost. It agrees with Theorem 1, where the cost gap decreases with $V$.

\begin{figure}[ht]
  \centering
  \includegraphics[width=0.75\hsize]{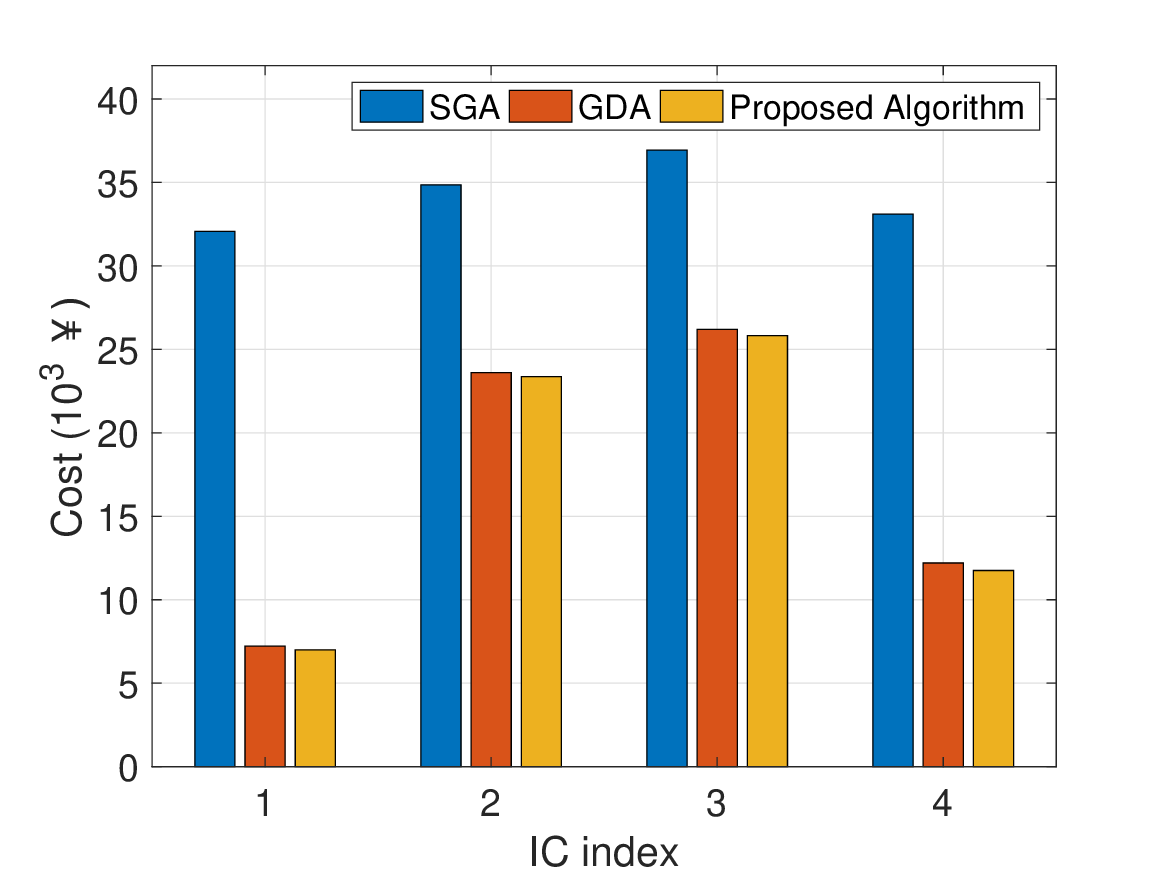}
  \caption{Total costs of four ICs under different methods.}
  \label{wqcc}
\end{figure}

\begin{figure}[ht]
  \centering
  \includegraphics[width=\hsize]{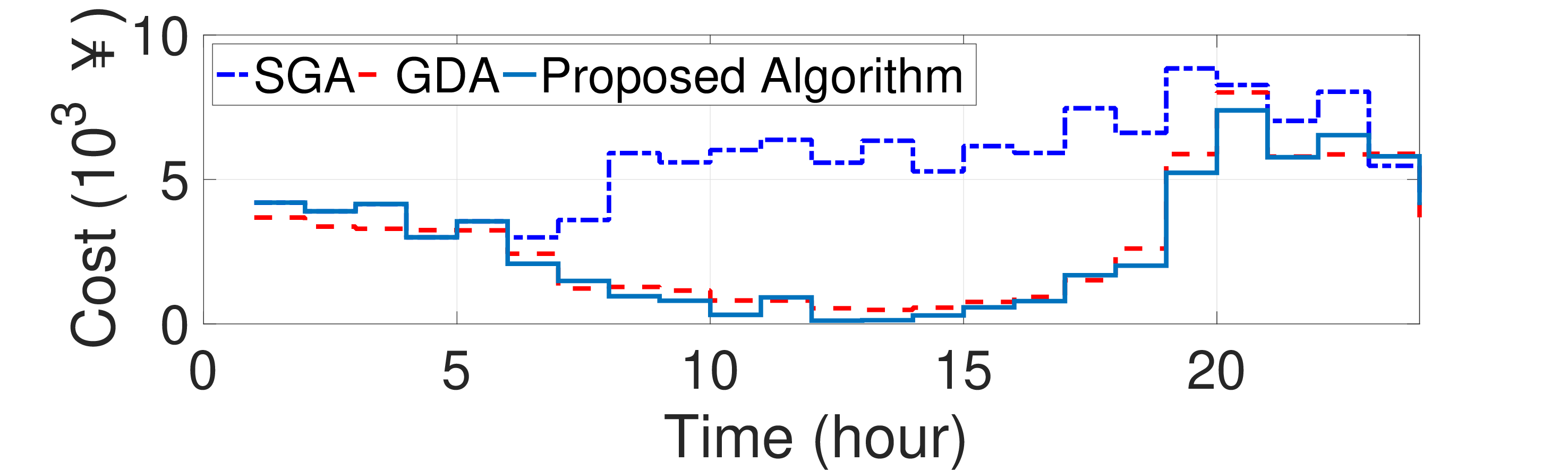}
  \caption{Cost comparison of four ICs under different methods.}
  \label{gdp24}
\end{figure}

 Fig.~\ref{wqcc} shows the total costs of four ICs under SGA, GDA and the proposed algorithm.  Fig.~\ref{gdp24} shows the cost comparison for four ICs at 24 hours under SGA, GDA and the proposed algorithm.  The overall cost under SGA approaches the overall cost under the proposed algorithm during 0:00-6:00, but more energy needs to be purchased during the day owing to lack of renewable energy, which causes higher cost. Since energy charging and discharging will not directly reduce the energy cost under GDA (i.e., without considering queue backlogs), the ICs cannot make full use of the battery and water tank, resulting in higher cost under GDA. In summary, the proposed algorithm can take full advantage of energy storage, minimize the time-average cost and ensure online energy scheduling.

\section{Conclusion}
The carbon emission and energy management problem of industrial clusters is investigated in this paper. A carbon and energy management framework is presented to achieve low-carbon energy scheduling, where a carbon flow model accompanying multi-energy flows is adopted to track and suppress carbon emissions on the user side. A bound tightening algorithm for constraints relaxation is adopted to deal with the quadratic constraint of gas flows. A carbon and energy trading model is constructed. The willingness to trade is improved by the ladder reward and punishment mechanism of carbon trading. A multi-energy trading and scheduling method combining Lyapunov optimization and matching game is presented, which further alleviates the imbalance of supply and demand and maximizes revenue. 
Finally, simulation results show that each industrial cluster reduces energy costs and carbon emissions, while relieving energy pressure. 

In this paper, four industrial clusters are considered in the industrial park. In some scenarios, a park may consist of more clusters. In addition, the energy efficiency and carbon emissions of industrial production can be improved by optimizing the production process. Therefore, joint trading and scheduling among more prosumers needs to be considered, such as {\cite{Ullah2022A}}, to significantly improve the social welfare. In addition, the industrial production process needs to be optimized, like {\cite{Li2022Integrated}}, to adapt to the energy and emission requirements.

\end{document}